\documentstyle[12pt,epsf,epsfig]{article}
\newcount\timecount
\newcount\hours \newcount\minutes  \newcount\temp \newcount\pmhours
\hours = \time
\divide\hours by 60
\temp = \hours
\multiply\temp by 60
\minutes = \time
\advance\minutes by -\temp
\def\hour{\the\hours}
\def\minute{\ifnum\minutes<10 0\the\minutes
            \else\the\minutes\fi}
\def\clock{
\ifnum\hours=0 12:\minute\ AM
\else\ifnum\hours<12 \hour:\minute\ AM
      \else\ifnum\hours=12 12:\minute\ PM
            \else\ifnum\hours>12
                 \pmhours=\hours
                 \advance\pmhours by -12
                 \the\pmhours:\minute\ PM
                 \fi
            \fi
      \fi
\fi
}

\def\monthname{\relax\ifcase\month 0/\or January\or February\or
   March\or April\or May\or June\or July\or August\or September\or
   October\or November\or December\else\number\month/\fi}

\def\bold#1{\setbox0=\hbox{$#1$}%
     \kern-.025em\copy0\kern-\wd0
     \kern.05em\copy0\kern-\wd0
     \kern-.025em\raise.0433em\box0 }

\def\beq{\begin{equation}}
\def\eeq{\end{equation}}

\def\ss{\scriptscriptstyle}
\def\ga{\mathrel{\raise.3ex\hbox{$>$\kern-.75em\lower1ex\hbox{$\sim$}}}}
\def\la{\mathrel{\raise.3ex\hbox{$<$\kern-.75em\lower1ex\hbox{$\sim$}}}}
\def\gev{{\rm \, Ge\kern-0.125em V}}
\def\tev{{\rm \, Te\kern-0.125em V}}
\def\gyr{{\rm \, G\kern-0.125em yr}}
\def\ohsq{\Omega_{\chi} h^2}

\def\nl{\hfill\nonumber\\&&}
\def\nnl{\hfill\nonumber\\}

\def\ttbt{\tan^2 \beta}

\def\gappeq{\mathrel{\rlap {\raise.5ex\hbox{$>$}}
{\lower.5ex\hbox{$\sim$}}}}
\def\lappeq{\mathrel{\rlap{\raise.5ex\hbox{$<$}}
{\lower.5ex\hbox{$\sim$}}}}
\def\Toprel#1\over#2{\mathrel{\mathop{#2}\limits^{#1}}}

\def\m12{m_{1\!/2}}

\def\mz{m_{\ss Z}}
\def\mA{m_{\ss A}}

\begin{document}
\begin{titlepage}
\pagestyle{empty}
\baselineskip=21pt
\vspace*{-0.6in}
\rightline{hep-ph/0302032}
\rightline{CERN--TH/2003-001}
\rightline{UMN--TH--2128/03}
\rightline{TPI--MINN--03/03}
\vskip 0.2in
\begin{center}
{\large{\bf Direct Detection of Dark Matter in the MSSM with Non-Universal 
Higgs Masses}}
\end{center}
\begin{center}
\vskip 0.2in
{{\bf John Ellis}$^1$, {\bf Andy Ferstl}$^{2}$, {\bf Keith
A.~Olive}$^{3}$ and {\bf Yudi Santoso}$^{3}$}\\
\vskip 0.1in
{\it
$^1${TH Division, CERN, Geneva, Switzerland}\\
$^2${Department of Physics, Winona State University, Winona, MN 55987, USA}\\
$^3${William I. Fine Theoretical Physics Institute, \\
University of Minnesota, Minneapolis, MN 55455, USA}}\\
{\bf Abstract}
\end{center}
\baselineskip=18pt \noindent

We calculate dark matter scattering rates in the minimal supersymmetric
extension of the Standard Model (MSSM), allowing the soft
supersymmetry-breaking masses of the Higgs multiplets, $m_{1,2}$, to be
non-universal (NUHM). Compared with the constrained MSSM (CMSSM) in which
$m_{1,2}$ are required to be equal to the soft supersymmetry-breaking
masses $m_0$ of the squark and slepton masses, we find that the
elastic scattering cross sections may be up to two orders
of magnitude larger than values in the CMSSM for similar LSP masses. We
find the following preferred ranges for the spin-independent cross
section:  $10^{-6}$~pb $\gappeq \sigma_{SI} \gappeq 10^{-10}$~pb, and for
the spin-dependent cross section: $10^{-3}$~pb $\gappeq \sigma_{SD}$, with
the lower bound on $\sigma_{SI}$ dependent on using the putative
constraint from the muon anomalous magnetic moment. We stress the
importance of incorporating accelerator and dark matter constraints in
restricting the NUHM parameter space, and also of requiring that no
undesirable vacuum appear below the GUT scale. In particular, values of
the spin-independent cross section another order of magnitude larger
would appear to be allowed, for small
$\tan \beta$, if the GUT vacuum stability requirement were relaxed,
and much lower cross-section values would be permitted if the muon
anomalous magnetic moment constraint were dropped.

\vfill
\leftline{CERN--TH/2003-001}
\leftline{January 2003}
\end{titlepage}
\baselineskip=18pt

\section{Introduction}

There have been many previous studies of the elastic scattering rates of
supersymmetric relic particles on matter in the minimal supersymmetric
extension of the Standard Model (MSSM) \cite{etal} - \cite{otherMSSMDM},
assuming conservation of
$R
\equiv (-1)^{3 B + L + 2 S}$, where $B$ is the baryon number, $L$ the
lepton number and $S$ the spin, so that the lightest supersymmetric
particle (LSP) is absolutely stable. As in most previous studies, we
assume this to be the lightest neutralino $\chi$~\cite{EHNOS}. In
this paper, we refine and extend previous calculations of the elastic
scattering rates when the input soft supersymmetry-breaking scalar masses
for the Higgs multiplets are allowed to be non-universal at the input GUT
scale, the non-universal Higgs model (NUHM).

As we discuss later in more detail, it is important to impose the
constraints due to accelerator experiments, including searches at LEP, $b
\to s \gamma$ and (optionally) the muon anomalous magnetic moment, $g_\mu 
- 2$. We assume also that most of the cold dark matter is
composed of LSPs, with relic density $0.1 < \ohsq < 0.3$, while being
aware that the lower part of this range currently appears the most
plausible \cite{MS}.

In the constrained MSSM (CMSSM), in which all the soft
supersymmetry-breaking scalar masses $m_0$ are assumed to be universal,
including those for the Higgs doublets $H_{1,2}$, the underlying
parameters may be taken as $m_0$, the soft supersymmetry-breaking gaugino
mass $m_{1/2}$ which is assumed to be universal, the trilinear
supersymmetry-breaking parameters $A_0$ that we set to zero at the GUT
scale in this paper, and the ratio $\tan \beta$ of Higgs vacuum
expectation values. In the NUHM \cite{oldnuhm} - \cite{Ellis:2002iu},
there are two additional free parameters, the two soft Higgs masses or
equivalently the Higgs superpotential coupling $\mu$ and the pseudoscalar
Higgs boson mass
$m_A$. These would be fixed by the electroweak symmetry-breaking vacuum
conditions in the CMSSM, up to a sign ambiguity in $\mu$, in terms of the
other parameters
$(m_0, m_{1/2}, A_0, \tan
\beta)$. We use the parameters $(m_0, m_{1/2}, \mu, m_A, A_0$, $\tan
\beta)$ to parametrize the more general NUHM.

As we have pointed out previously \cite{Ellis:2002wv,Ellis:2002iu}, this
six-dimensional NUHM parameter space is significantly restricted by the
requirement that no undesirable vacuum appears when one uses the
renormalization-group equations to run the soft supersymmetry-breaking
parameters between the input GUT scale and the electroweak scale. This
requirement constrains the non-universalities of the Higgs masses: ${\hat
m}_i \equiv {\rm sign}(m_i^2) |{m_i / m_0}|: i = 1,2$, which in turn
restricts the range of elastic scattering cross sections that we
find~\footnote{This extended stability requirement would also exclude
non-universalities for the input squark and slepton masses that allowed
their physical values to be similar, an assumption that might lead to
much larger elastic scattering cross sections than we find here.}.

The allowed regions in the $(\mu, \mA)$, $(\mu, M_2)$ and $(m_{1/2}, m_0)$
planes for certain discrete values of the other NUHM parameters have been
described in~\cite{Ellis:2002iu}. Our first step in this paper is to
provide contours of the elastic scattering cross sections in selected
planes, providing a first comparison with the CMSSM points that appear in
these planes. Secondly, we display the ranges of elastic scattering cross
sections that are allowed in these planes, as functions of the LSP mass.  
In general, we find that the spin-independent elastic scattering cross 
sections may be up to two orders of magnitude
larger than values in the CMSSM for similar LSP masses, and another order 
of magnitude larger if the GUT vacuum stability requirement is relaxed.
Thirdly, we display ranges of the elastic scattering 
cross sections as functions of the LSP mass for all allowed values of 
the other NUHM parameters.

We review the NUHM in Section 2, including the experimental and
phenomenological constraints on its parameter space. Then, in Section 3,
we summarize our treatment of the elastic scattering matrix elements and
display contours of the cross sections in various planar projections of
the NUHM parameter space.  Section 4 presents and discusses the ranges
of the cross sections attainable in the NUHM. Finally, Section 5
draws some conclusions from our analysis.

\section{The NUHM and Constraints on its Parameter Space}

We assume that the soft supersymmetry-breaking parameters are specified at
some large input scale $M_X$, such as the supergravity or grand
unification scale. Motivated by restrictions on flavour-changing neutral
interactions, we assume that squarks and sleptons with the same Standard
Model quantum numbers have universal soft supersymmetry-breaking scalar
masses at this input scale. With the weaker justification provided by some
GUTs, we further assume universality between the soft scalar masses of
squarks and sleptons. However, in the NUHM studied here, we allow the soft
supersymmetry-breaking scalar contributions to the masses of the Higgs
supermultiplets at $M_X$ to be free non-universal parameters. Their
running from $M_X$ down to low energies relates $m^2_1(M_X)$ and
$m^2_2(M_X)$ to the Higgs supermultiplet mixing parameter $\mu$ and the
pseudoscalar Higgs mass $m_A$. Therefore, we use as free parameters
$\mu(\mz) \equiv \mu$ and $m_A(Q) \equiv m_A$, where $Q\equiv
(m_{\widetilde{t}_R} m_{\widetilde{t}_L})^{1/2}$, in addition to the
parameters $(m_0(M_X), m_{1/2}(M_X), A_0, \tan \beta)$ used in the
CMSSM~\footnote{In this paper, we use the input $A_0 = 0$ for 
definiteness, noting that the range of effective low-energy values of 
$A$, after renormalization below the GUT scale, is quite limited.}.
 
The electroweak vacuum conditions may be written in the form:
\begin{equation}
m_A^2 (Q) = m_1^2(Q) + m_2^2(Q) + 2 \mu^2(Q) + \Delta_A(Q)
\end{equation}
and
\begin{equation}
\mu^2 = \frac{m_1^2 - m_2^2 \tan^2 \beta + \frac{1}{2} \mz^2 (1 - \tan^2 \beta)
+ \Delta_\mu^{(1)}}{\tan^2 \beta - 1 + \Delta_\mu^{(2)}},
\end{equation}
where $\Delta_A$ and $\Delta_\mu^{(1,2)}$ are loop
corrections~\cite{Barger:1993gh,deBoer:1994he,Carena:2001fw} and $m_{1,2} \equiv
m_{1,2}(\mz)$.  
We incorporate the known radiative
corrections~\cite{Barger:1993gh,IL,Martin:1993zk} 
$c_1, c_2$ and $c_\mu$ relating the values of the
NUHM parameters at $Q$ to their values at $\mz$: 
\begin{eqnarray}
m_1^2(Q) &=& m_1^2 + c_1 \nnl
m_2^2(Q) &=& m_2^2 + c_2 \nnl
\mu^2(Q) &=& \mu^2 + c_\mu \, .
\end{eqnarray}
Solving for $m^2_1$ and $m^2_2$, one has
\begin{eqnarray}
m_1^2(1+ \tan^2 \beta) &=& m_A^2(Q) \tan^2 \beta - \mu^2 (\tan^2 \beta + 1 -
\Delta_\mu^{(2)} ) 
- (c_1 + c_2 + 2 c_\mu) \ttbt \nl - \Delta_A(Q) \ttbt 
- \frac{1}{2} \mz^2 (1 - \ttbt) - \Delta_\mu^{(1)} 
\label{m1}
\end{eqnarray}
and 
\begin{eqnarray}
m_2^2(1+ \tan^2 \beta) &=& m_A^2(Q) - \mu^2 (\tan^2 \beta + 1 +
\Delta_\mu^{(2)} )
- (c_1 + c_2 + 2 c_\mu) \nl
- \Delta_A(Q) + \frac{1}{2} \mz^2 (1 - \ttbt) + \Delta_\mu^{(1)},
\label{m2}
\end{eqnarray}
which we use to perform our numerical calculations. 

It can be seen from (\ref{m1}) and (\ref{m2}) that, if $m_A$ is too small
or $\mu$ is too large, then $m_1^2$ and/or $m_2^2$ can become negative and
large. This could lead to $m_1^2(M_X) + \mu^2(M_X) < 0$ and/or $m_2^2(M_X)
+ \mu^2(M_X) < 0$, thus triggering electroweak symmetry breaking at the
GUT scale. The requirement that electroweak symmetry breaking occurs far
below the GUT scale forces us to impose the conditions $m_1^2(M_X)+
\mu(M_X), m_2^2(M_X)+ \mu(M_X) > 0$ as extra constraints, which we call
the GUT stability constraint~\footnote{For a different point of view,
however, see~\cite{fors}.}. We have discussed in~\cite{Ellis:2002iu}
issues related to the NUHM renormalization group equations (RGE's) and
their solutions.

We assume that $R$ parity is conserved, so that the LSP is stable and is
present in the Universe today as a relic from the Big Bang. Searches for
anomalous heavy isotopes tell us that the dark matter should be
weakly-interacting and neutral, and therefore eliminate all but the
neutralino and the sneutrinos as possible LSPs. LEP and direct dark-matter
searches together exclude a sneutrino LSP~\cite{Falk:1994es}, at least if
the majority of the CDM is the LSP. Thus we require in our analysis that
the lightest neutralino be the LSP. We include in our analysis all 
relevant coannihilation processes involving this LSP and sparticles that 
may become near-degenerate in various regions of the NUHM parameter space. 
We restrict our attention to regions of the NUHM parameter space where 
$0.1 < \ohsq < 0.3$.

We impose in our analysis the constraints provided by direct sparticle
searches at LEP, including that on the lightest chargino $\chi^\pm$:
$m_{\chi^\pm} \gappeq$ 103.5 GeV~\cite{LEPsusy}, and that on the selectron
$\tilde e$: $m_{\tilde e}\gappeq$ 99 GeV \cite{LEPSUSYWG_0101}. Another
important constraint is provided by the LEP lower limit on the Higgs mass:
$m_H > 114.4$~GeV \cite{LEPHiggs} in the Standard Model\footnote{In view
of the theoretical uncertainty in calculating $m_h$, we apply this bound
with just three significant digits, i.e., our figures use the constraint 
$m_h >
114$ GeV.}. The lightest Higgs boson $h$ in the general MSSM must obey a
similar limit, which may in principle be relaxed for larger $\tan \beta$.
However, as we discussed in our previous analysis of the
NUHM~\cite{Ellis:2002wv}, the relaxation in the LEP limit is not relevant
in the regions of MSSM parameter space of interest to us. We recall that
$m_h$ is sensitive to sparticle masses, particularly $m_{\tilde t}$, via
loop corrections~\cite{radcorrH,FeynHiggs}, implying that the LEP Higgs
limit constrains the NUHM parameters. We also impose the constraint
imposed by measurements of $b\rightarrow s\gamma$~\cite{bsg}, as discussed
in~\cite{Ellis:2002iu}.

We take an agnostic attitude towards the latest value of the anomalous
magnetic moment of the muon reported~\cite{newBNL} by the BNL E821
experiment. The world average of $a_\mu\equiv {1\over 2} (g_\mu -2)$ now
deviates by $(33.7 \pm 11.2) \times 10^{-10}$ from the Standard Model
calculation of~\cite{Davier} using $e^+ e^-$ data, and by $(9.4 \pm 10.5)
\times 10^{-10}$ from the Standard Model calculation of~\cite{Davier}
based on $\tau$ decay data.  On some of the subsequent plots, we display
the formal 2-$\sigma$ range $11.3 \times 10^{-10} < \delta a_\mu < 56.1
\times 10^{-10}$. However, in view of the chequered history
of the theoretical Standard Model calculations of $a_\mu$, we do not impose
this as an
absolute constraint on the supersymmetric parameter space.

The results of applying the above constraints to various two-dimensional 
projections of the NUHM parameter space were described
in~\cite{Ellis:2002iu}.

\section{Contours of the Cross Sections for Elastic Scattering}

The code we use to calculate the spin-independent and -dependent elastic
dark matter scattering cross sections $\sigma_{SI,SD}$ was documented
in~\cite{EFlO1,EFlO2}, together with the ranges of values of the hadronic
matrix elements that we use. The cross sections for protons and neutrons
are similar within the quoted uncertainties in these matrix elements.
There are other codes available~\cite{NeutDriver} that include additional
contributions to the scattering matrix elements, but a
comparison~\cite{EFFMO} showed that the improvements were not essential
for the CMSSM, and we believe they may also be neglected for our
comparisons of the NUHM.

In~\cite{Ellis:2002iu}, we analyzed NUHM dark matter in two ways:  (i)
fixing
$\tan \beta = 10$ and $\mu > 0$, but choosing different values of $\mu$
and $m_A$, rather than assuming the CMSSM values, and (ii) varying $\tan
\beta$ for representative fixed values of $\mu$ and $m_A$. We presented
in~\cite{Ellis:2002iu} three types of slices through the NUHM parameter
space, along $(m_{1/2}, m_0)$ planes, $(\mu, m_A)$ planes and $(\mu, M_2)$
planes.

In this paper, we concentrate first on a few specific examples of these
slices, presenting later more general results.  We choose two
representative examples each of the $(m_{1/2}, m_0)$ planes, $(\mu, m_A)$
planes and $(\mu, M_2)$ planes shown previously. As we discuss later in
more detail, the dependences of the cross sections on $\tan \beta$ are
weaker than those on some other parameters, so we concentrate on planes
for $\tan \beta = 10$. The examples we choose are the $(m_{1/2}, m_0)$
planes for $\mu = 400~{\rm GeV}$, $m_A = 400$~GeV and $\mu = 700~{\rm GeV}$,
$m_A = 700$~GeV, corresponding to Figs.~2(a) and (d)
of~\cite{Ellis:2002iu}, the
$(\mu, m_A)$ planes for $m_0 = 100~{\rm GeV}$, $m_{1/2} = 300$~GeV and $m_0
= 300~{\rm GeV}$, $m_{1/2} = 300$~GeV, corresponding to Figs.~4(a) and (c)
of~\cite{Ellis:2002iu}, and the $(\mu, M_2)$ planes for $m_0 = 100~{\rm
GeV}$, $m_A = 300$~GeV and $m_0 = 300~{\rm GeV}$, $m_A = 500$~GeV,
corresponding to Figs.~8(a) and (c) of~\cite{Ellis:2002iu}.

\subsection{Examples of $(m_{1/2}, m_0)$ Planes}

\begin{figure}
\vspace*{-0.75in}
\begin{minipage}{8in}
\epsfig{file=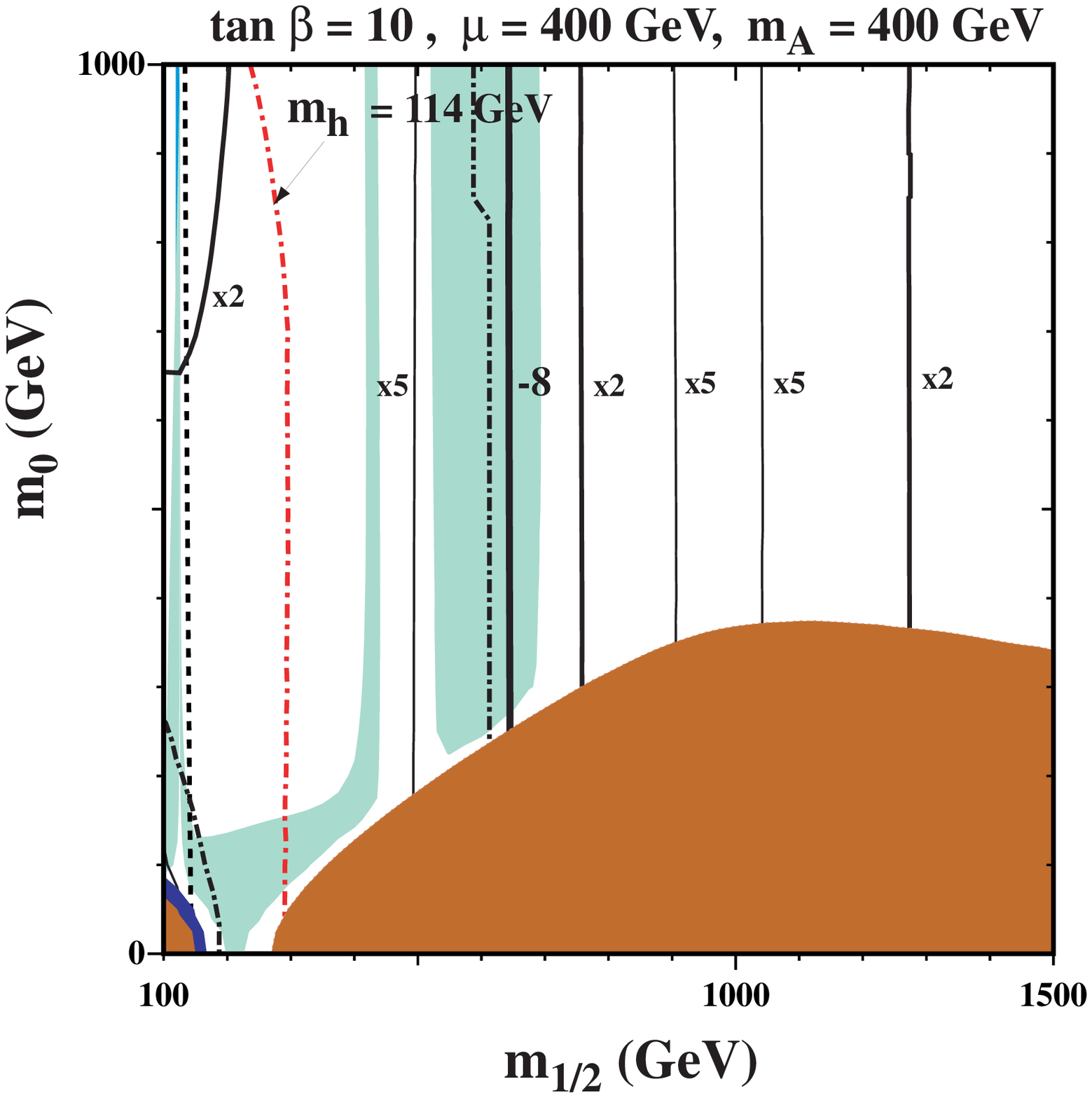,height=3.2in}
\epsfig{file=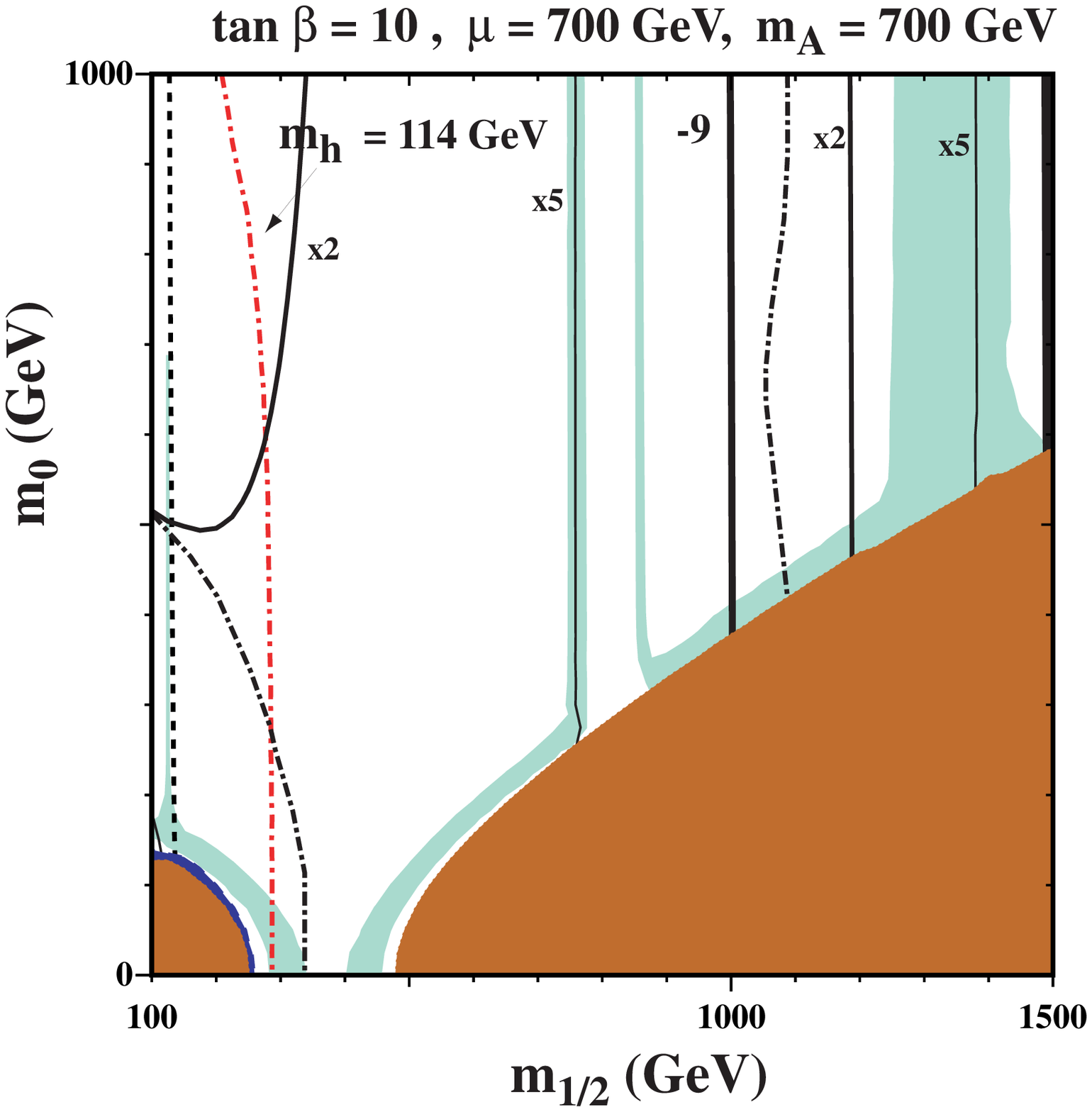,height=3.2in} \hfill
\end{minipage}
\begin{minipage}{8in}
\epsfig{file=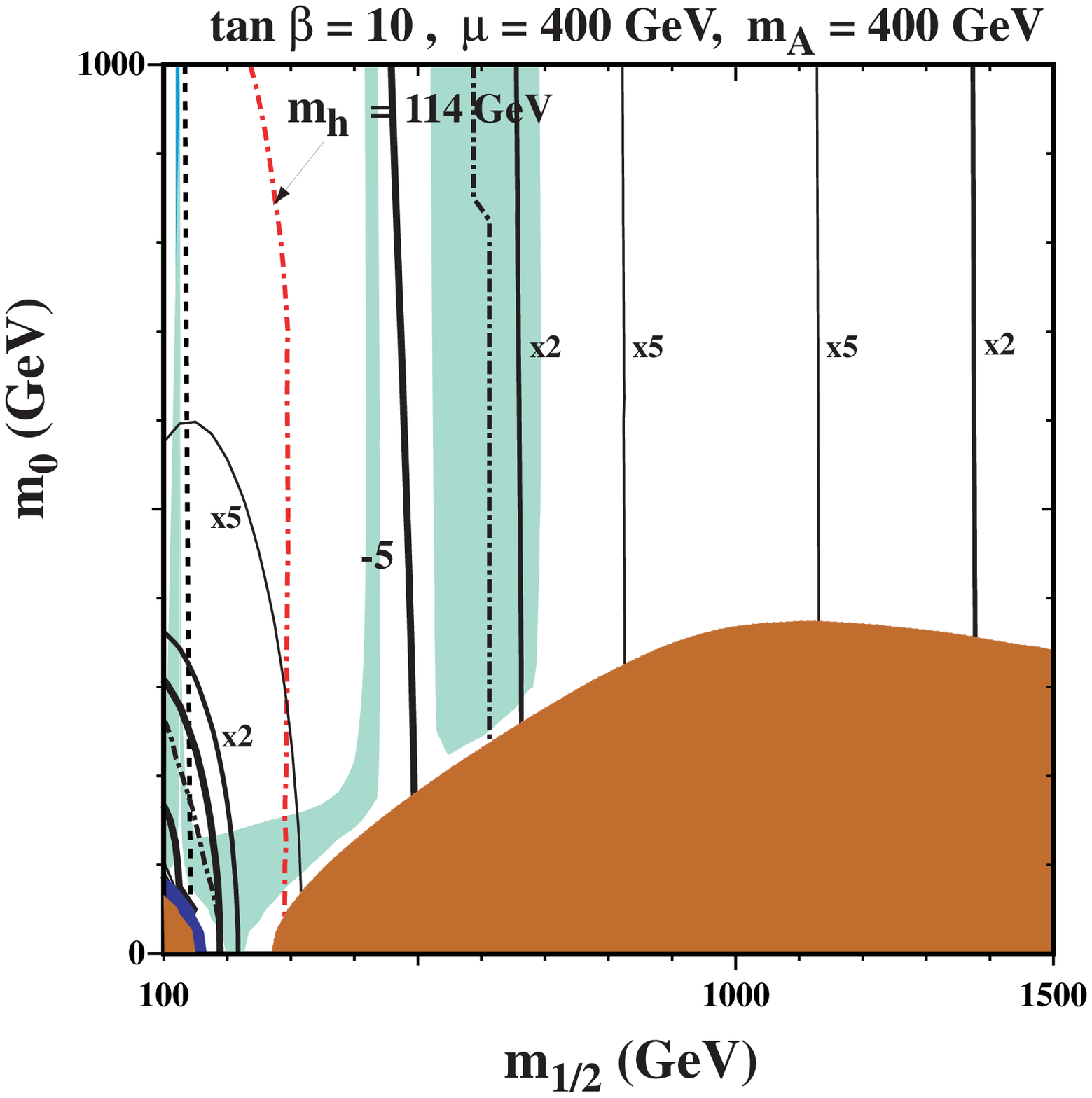,height=3.2in}
\epsfig{file=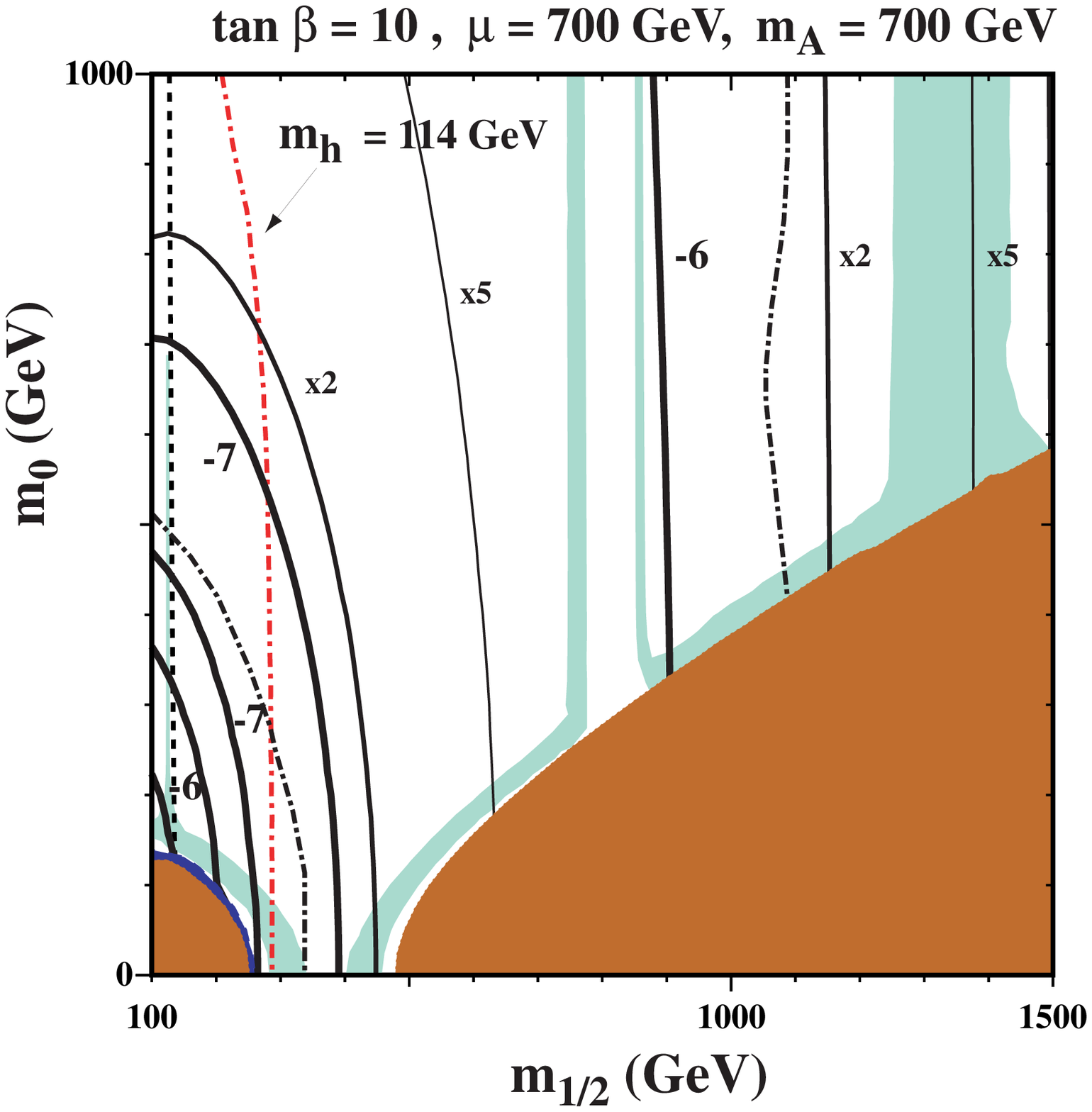,height=3.2in} \hfill
\end{minipage}
\caption{\label{fig:contourmm}
{\it 
Contours of (a,b) the spin-independent and (c,d) the spin-dependent cross
sections (solid black lines) are superimposed on the NUHM $(m_{1/2}, m_0)$
planes for $\tan \beta = 10$ and (a,c) $\mu = 400$~GeV and $\mA =
400$~GeV, and (b,d) $\mu = 700$~GeV and $\mA = 700$~GeV. The near-vertical
(red) dot-dashed lines are the contours $m_h = 114$~GeV, and the
near-vertical (black) dashed lines at lower $m_{1/2}$ are the contours
$m_{\chi^\pm} = 103.5$~GeV.  The dark (brick red) shaded regions is
excluded because a charged particle is lighter than the neutralino, and
the darker (dark blue) shaded regions is excluded because the LSP is a
sneutrino.  The light (turquoise) shaded areas are the cosmologically
preferred regions with \protect\mbox{$0.1\leq\ohsq\leq 0.3$}.  The dark
(black) dot-dashed lines indicate the GUT stability constraint. There are
two such lines for each panel and only the areas in between are allowed by
this constraint.  }}
\end{figure}

We display in Fig.~\ref{fig:contourmm} contours of (a,b) the
spin-independent and (c,d) the spin-dependent elastic scattering cross
sections, in the cases $\tan \beta = 10$, and (a,c) $\mu = 400$~GeV and
$\mA = 400$~GeV, and (b,d) $\mu = 700$~GeV and $\mA = 700$~GeV. We assume
here and in the subsequent figures that $A_0 = 0$, $m_t = 175$~GeV and
$m_b(m_b)^{\overline {MS}}_{SM} = 4.25$~GeV. Here and elsewhere, the
thickest contours denote decades in the cross-section values in pb,
labelled by their exponents. The medium and thinnest lines are
intermediate contours in the cross-section values, namely $2 \times$ and
$5 \times$ decades, as labelled.

We notice immediately that the cross-section contours are nearly vertical
at large $m_{1/2}$, reflecting the fact that they become almost
independent of $m_0$ in the NUHM. We also notice that, within the GUT
stability range (inside the black dot-dashed curves), the cross sections
{\it increase} with
$m_{1/2}$. This is because the LSP becomes more Higgsino-like as
$m_{1/2}$ increases. However, the cross sections do decrease again for
very large $m_{1/2}$ beyond the GUT stability limit as the low energy
scalar masses increase with $m_{1/2}$ as does the light Higgs mass
(though slowly).

The GUT stability requirement
imposes $m_{1/2} \lappeq 600 (1100)$~GeV in panels (a,c) and (b,d),
respectively. Because of the increases in the cross sections with
$m_{1/2}$, there are in turn {\it upper bounds} on the cross sections,
that would not be respected if GUT stability were disregarded. Because the
$g_\mu - 2$ constraint would provide even stronger upper bounds on
$m_{1/2}$, it would also impose stronger upper bounds on the cross
sections. In Figs.~\ref{fig:contourmm} - \ref{fig:contourmuM2}, we have
left off the $g_\mu -2$ contours to avoid confusion with the cross section
contours
we are highlighting here. For the case with $\mu = 400$ GeV and $m_A =
400$ GeV, the $g_\mu -2$ constraint places an upper limit on $m_{1/2} $ of
about 400 GeV and for the case with $\mu = 700$ GeV and $m_A =
700$ GeV, the limit is $m_{1/2} \la 450 $GeV. 

As already remarked, the cross sections themselves do not vary greatly
with $\tan \beta$, but the interplay of the other constraints is rather
$\tan \beta$-dependent. In particular, at large $\tan \beta$ the $g_\mu -
2$ constraint would not reduce significantly the upper bounds on the cross
sections.

\subsection{Examples of $(\mu, m_A)$ Planes}

\begin{figure}
\vspace*{-0.75in}
\begin{minipage}{8in}
\epsfig{file=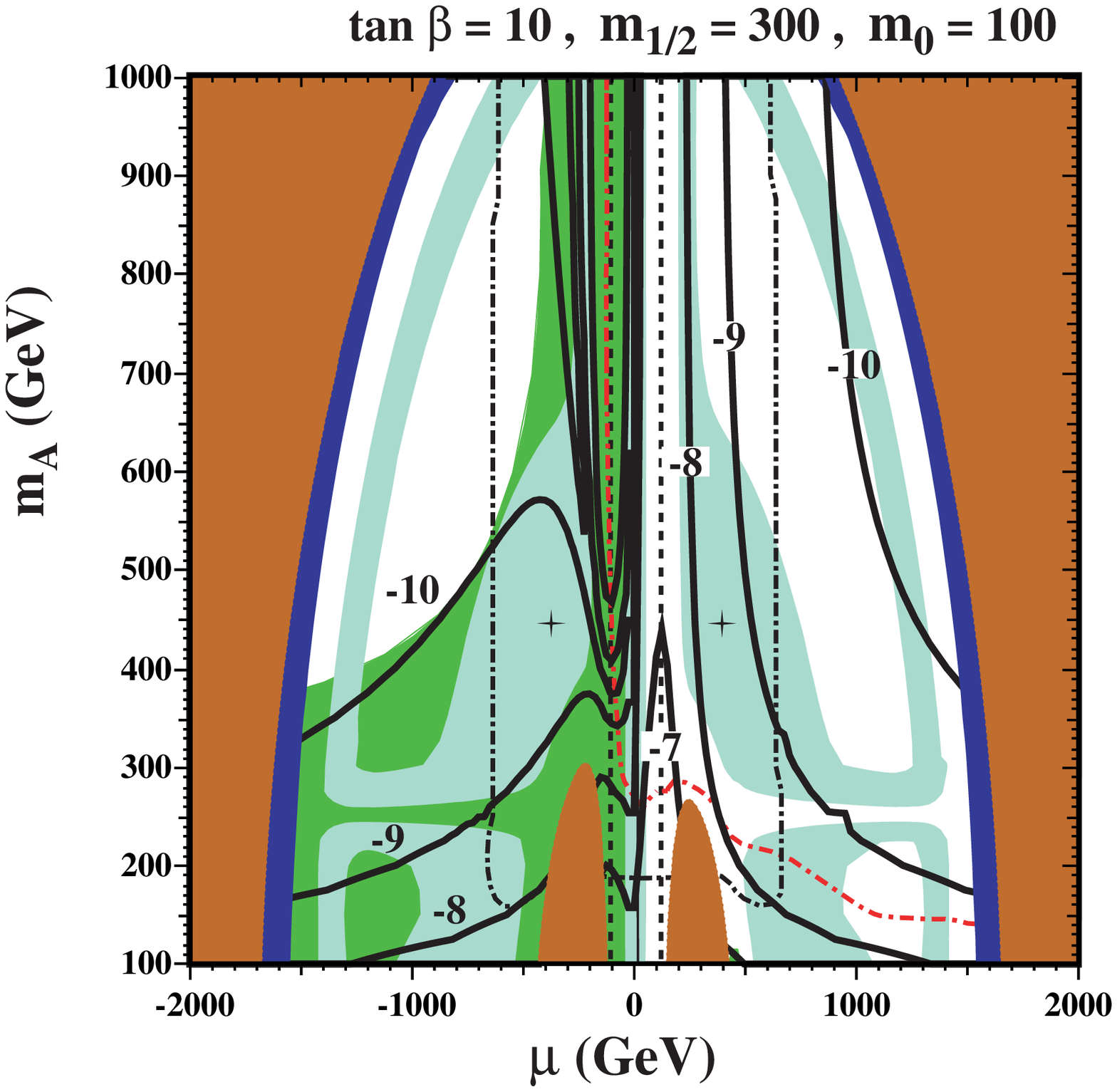,height=3.2in}
\epsfig{file=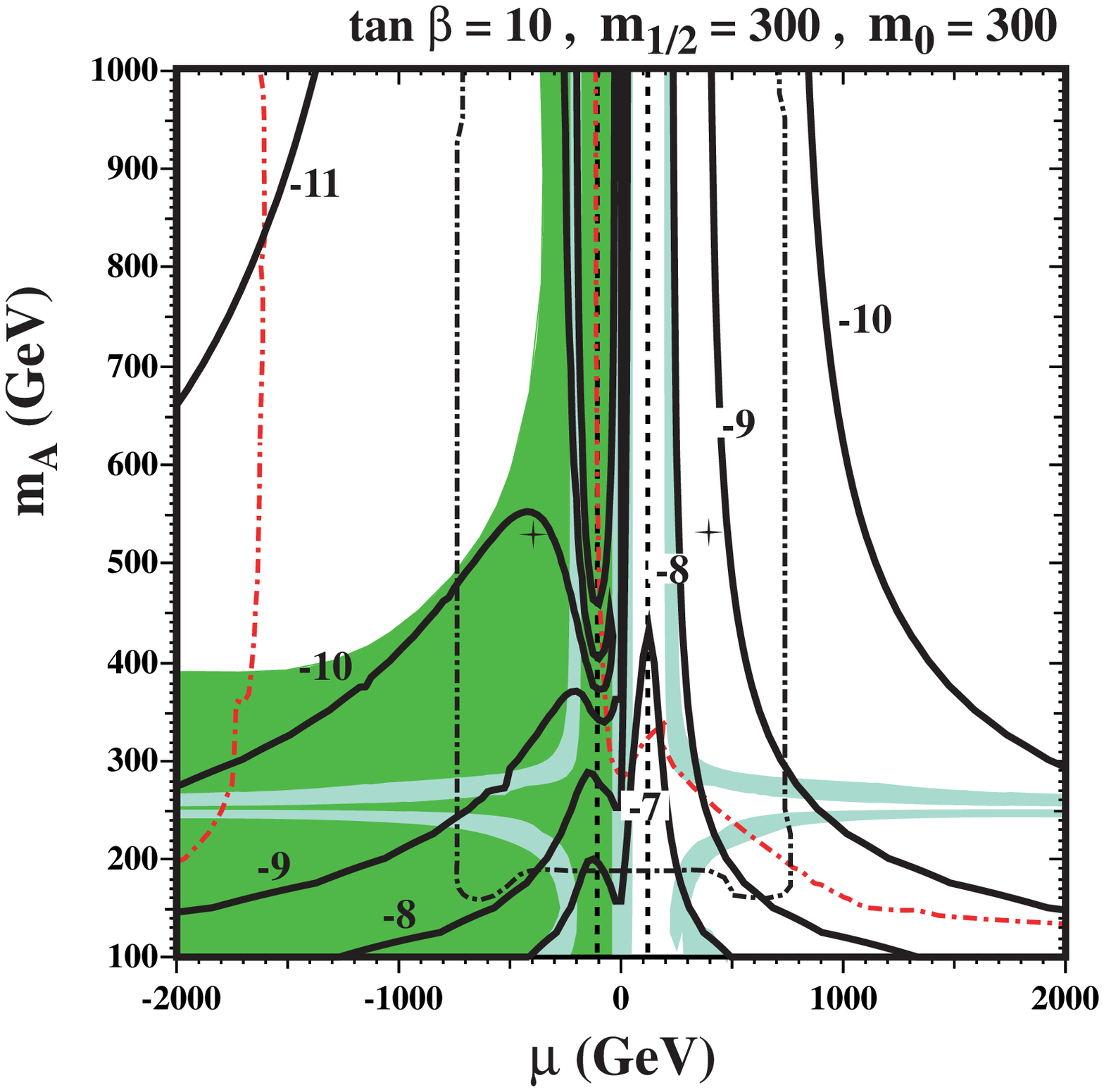,height=3.2in} \hfill
\end{minipage}
\begin{minipage}{8in}
\epsfig{file=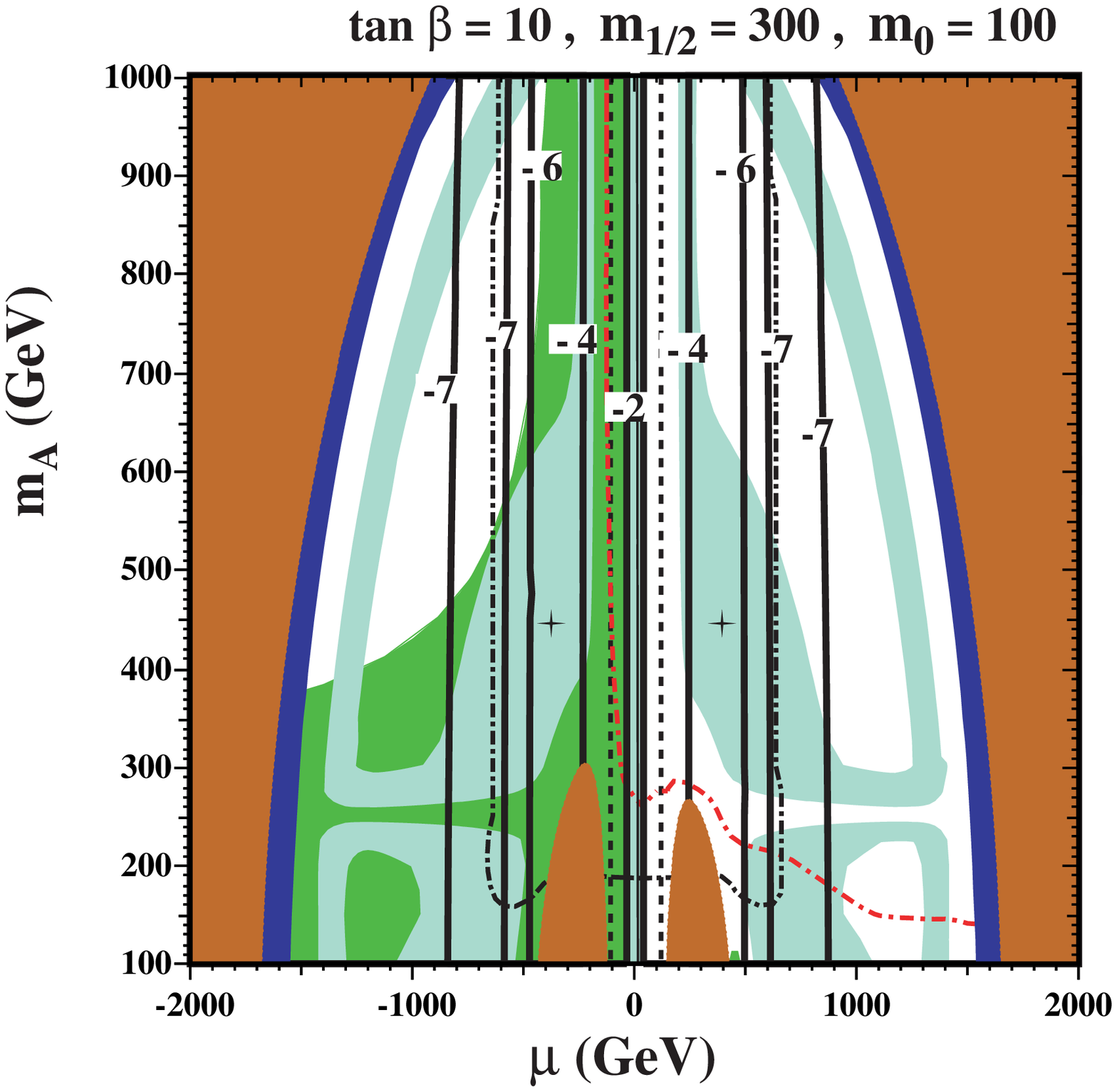,height=3.2in}
\epsfig{file=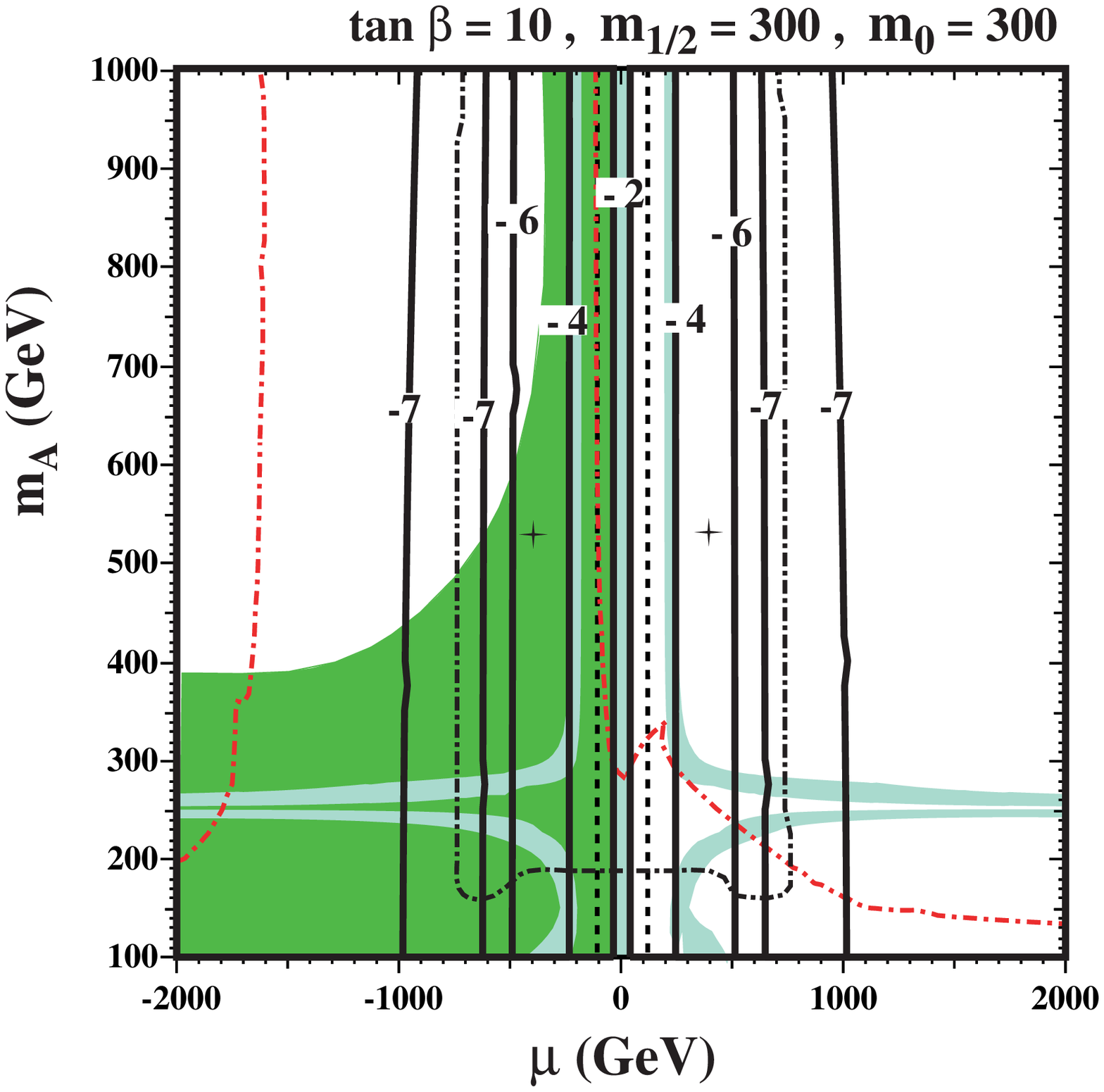,height=3.2in} \hfill
\end{minipage}
\caption{\label{fig:contourmumA}
{\it 
Contours of (a,b) the spin-independent and (c,d) the spin-dependent cross
sections (solid black lines) are superimposed on the NUHM $(\mu, \mA)$ 
planes for $\tan \beta = 10$, (a,c) $m_0 = 100$~GeV 
and $m_{1/2} = 300$~GeV, (b,d) $m_0 =
300$~GeV and $m_{1/2} = 300$~GeV. The shadings and
line styles are the same as in Fig.~\ref{fig:contourmm}. Here, we see
in addition the constraint from $b \to s \gamma$.  The excluded region
is medium (green) shaded. The crosses  denote the CMSSM points for these
choices of $m_0$ and $m_{1/2}$.}}
\end{figure}

We display in Fig.~\ref{fig:contourmumA} contours of (a,b) the
spin-independent and (c,d) the spin-dependent elastic scattering cross
sections, in the cases $\tan \beta = 10$, and (a,c) $m_0 = 100$~GeV and
$m_{1/2} = 300$~GeV, (b,d) $m_0 = 300$~GeV and $m_{1/2} = 300$~GeV. We see
that there are large suppressions in the spin-independent cross section
for $\mu \sim - 100$~GeV and $m_A \gappeq 500$~GeV, reflecting a
cancellation in the matrix element. Apart from this, the cross sections
generally decrease with increasing $|\mu|$ and (to a lesser extent)
$m_A$. In this sense, the lower bounds on $|\mu|$ and $m_A$ set an upper bound
on the cross section in the allowed region.

In the CMSSM, cancellations which drive the cross section to extremely
small values occur only in the spin-independent case, and only for $\mu <
0$.  In the NUHM model, however, we find that there is a new source for a
cancellation which affects the spin-dependent cross section for either
sign of $\mu$. The reason this occurs is as follows.  In the CMSSM, the
spin-dependent cross section is dominated by the up-squark exchange term
and, despite the difference in the relative signs of the up-type and
down-type contributions, the total cross section remains reasonably large.
On the other hand, in the NUHM case considered here, low-energy sfermion
masses are affected by the splitting between the soft
supersymmetry-breaking Higgs masses at the GUT scale, $S = g_1^2 (m_1^2 -
m_2^2)/4$. When $S \ne 0$, the up squarks get somewhat heavier and the
down squarks somewhat lighter as $S$ increases (see~\cite{Ellis:2002iu}
for details on the effects of $S$ in the renormalization-group equations),
opening up the possibility of a cancellation between the contributions.  
In Fig.~\ref{fig:contourmumA}(c,d), the cancellation occurs between the
two $10^{-7}$ pb contours very close to, but outside the GUT stability
curve. The GUT stability requirement, that bounds $| \mu | \lappeq
700$~GeV in the figures displayed, therefore provides {\it lower bounds}
on the spin-dependent cross sections. These are somewhat lower than the
values found at the CMSSM points indicated by crosses in
Fig.~\ref{fig:contourmumA}.  The same cannot be said for the
spin-independent cross-sections, because the very small cross sections due
to cancellations occur within the GUT stability region, except when one
applies the $g_\mu -2$ constraint, which excludes the $\mu <0$ region. The
interplay of the other constraints is more complicated: as usual, $\mu <
0$ is disfavoured by the $m_h$, $b \to s \gamma$ and $g_\mu - 2$
constraints. The upshot for $\mu > 0$ is that the cross sections are
bounded above by the $\ohsq$ constraint, so that they cannot be much more
than a factor of $\sim 10$ greater than the CMSSM values.

For the values of parameters chosen in Fig.~\ref{fig:contourmumA} (a,c),
we obtain a spin-independent cross-section of $\sigma_{SI}  = 2.6 \times
10^{-9}$ pb and a spin-dependent cross-section of $\sigma_{SD} = 5.4
\times 10^{-6}$ pb in the CMSSM for $\mu >0$ (cf. the position of the crosses
in the figures).  In the NUHM, we find that the range of possible cross sections
(for this case) is $3 \times
10^{-10} {\rm pb} \la \sigma_{SI} \la  3 \times 10^{-8} {\rm pb}$
and $3 \times
10^{-8} {\rm pb} \la \sigma_{SD} \la  1.6 \times 10^{-4} {\rm pb}$
when all constraints other than $g_\mu -2$ are included.
 For the parameters in
Fig.~\ref{fig:contourmumA} (b,d), we find the CMSSM spin-independent
cross section $\sigma_{SI}  = 1.9 \times 10^{-9}$ pb, while the
spin-dependent cross section is relatively unchanged. Note, however, that
this CMSSM point would be excluded due to an excessive value for $\Omega
h^2$ ( $> 1$). In the NUHM, this parameter choice is allowed and gives
the range  
$ 10^{-9} {\rm pb} \la \sigma_{SI} \la  8 \times 10^{-8} {\rm pb}$
and $9 \times
10^{-9} {\rm pb} \la \sigma_{SD} \la  2 \times 10^{-4} {\rm pb}$
for the elastic cross sections. In this case, there is essentially no
`bulk' cosmological region, and the spread in $\sigma$ is due
to the region where the relic density is due to the heavy Higgs 
$s$-channel exchange, allowing for a larger range in $\mu$.

\subsection{Examples of $(\mu, M_2)$ Planes}

\begin{figure}
\vspace*{-0.75in}
\begin{minipage}{8in}
\epsfig{file=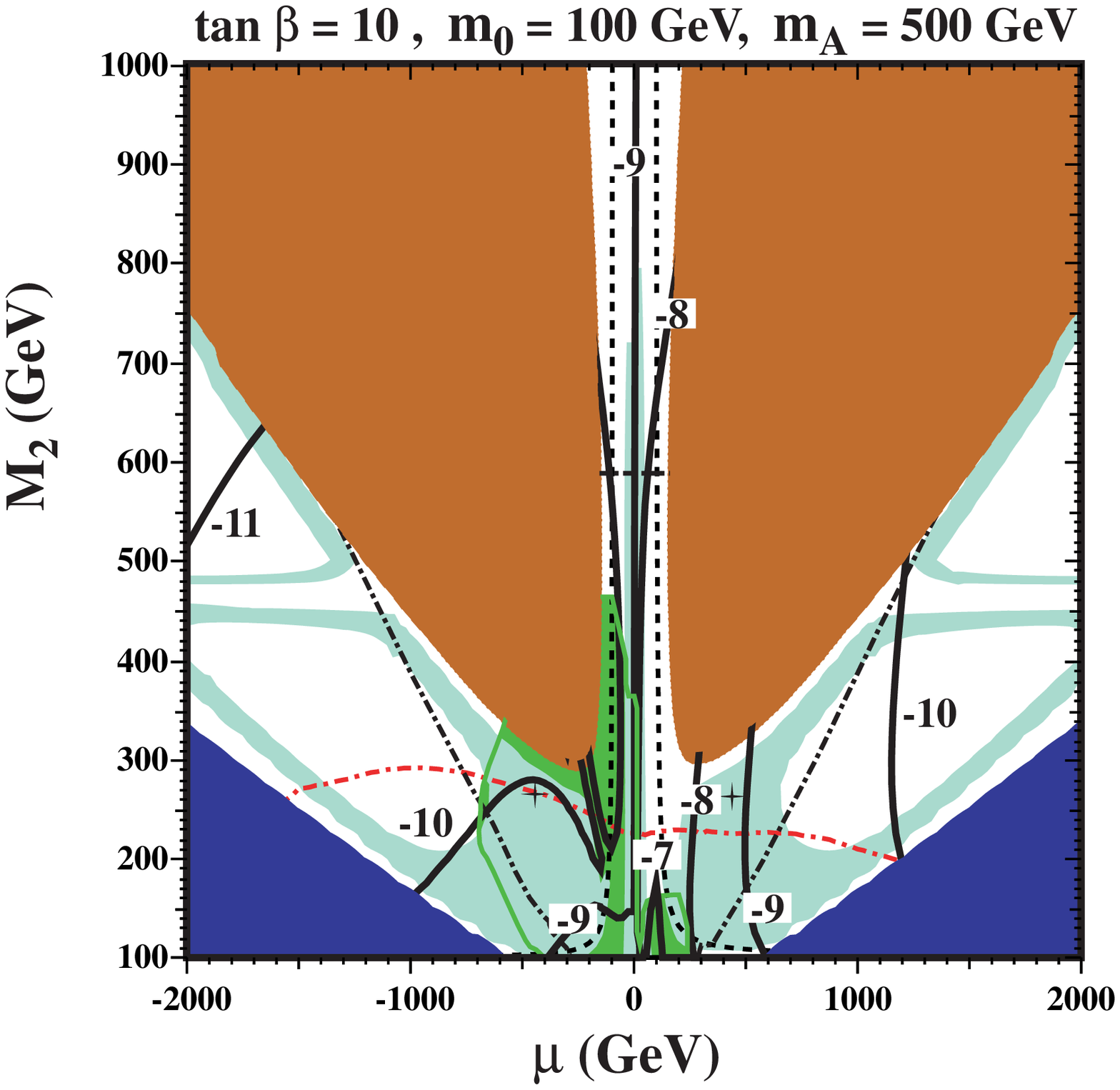,height=3.2in}
\epsfig{file=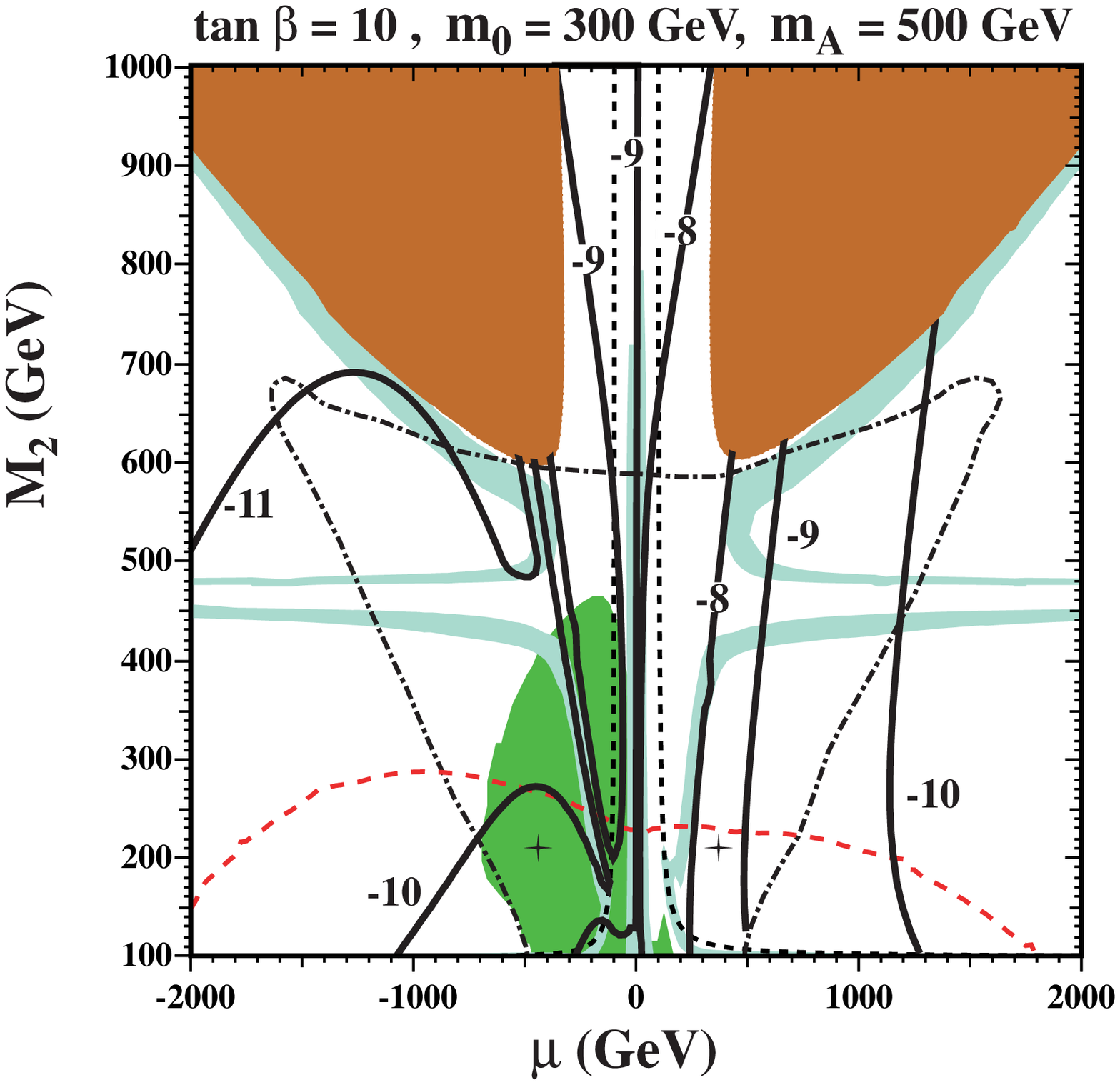,height=3.2in} \hfill
\end{minipage}
\begin{minipage}{8in}
\epsfig{file=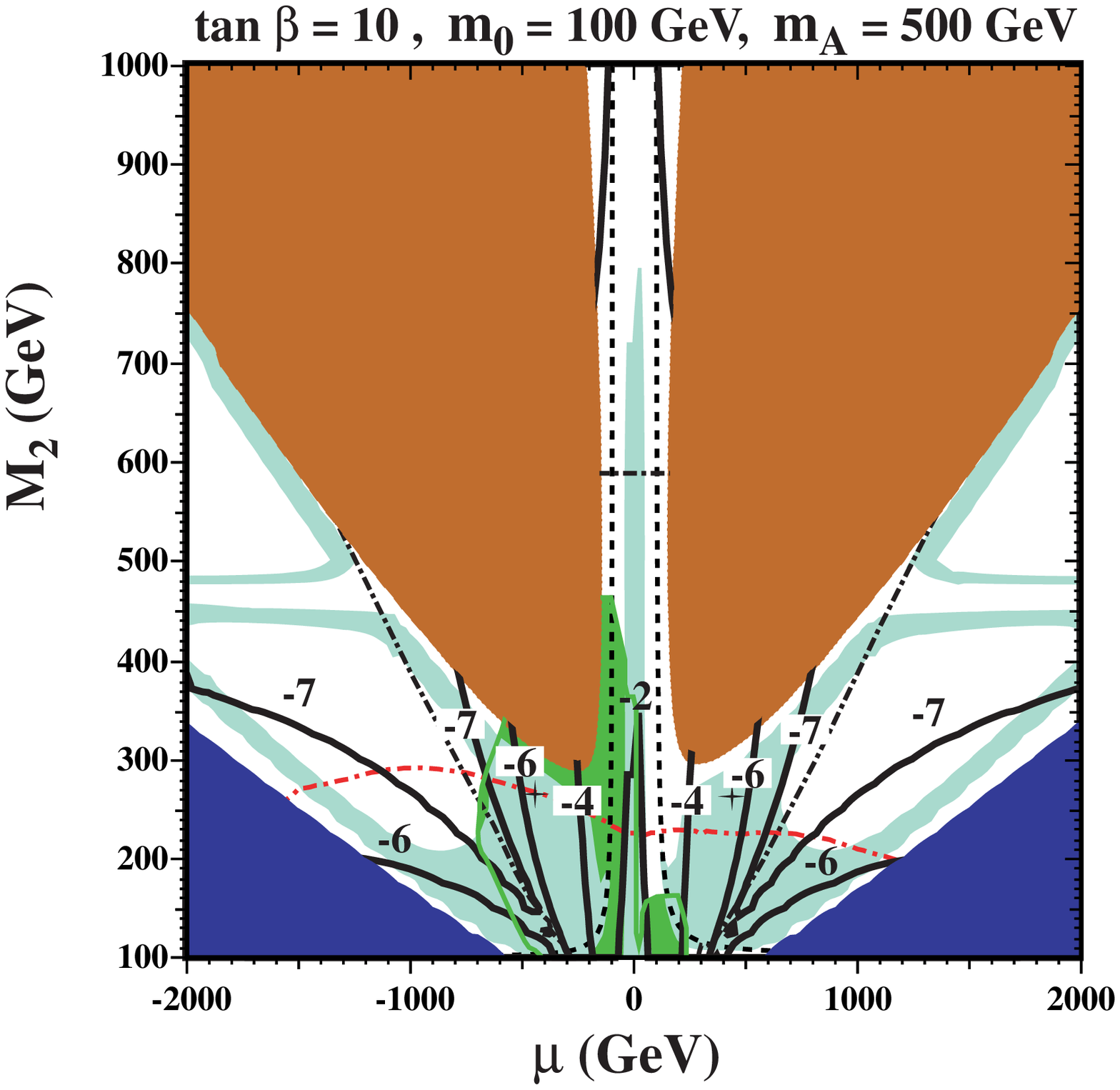,height=3.2in}
\epsfig{file=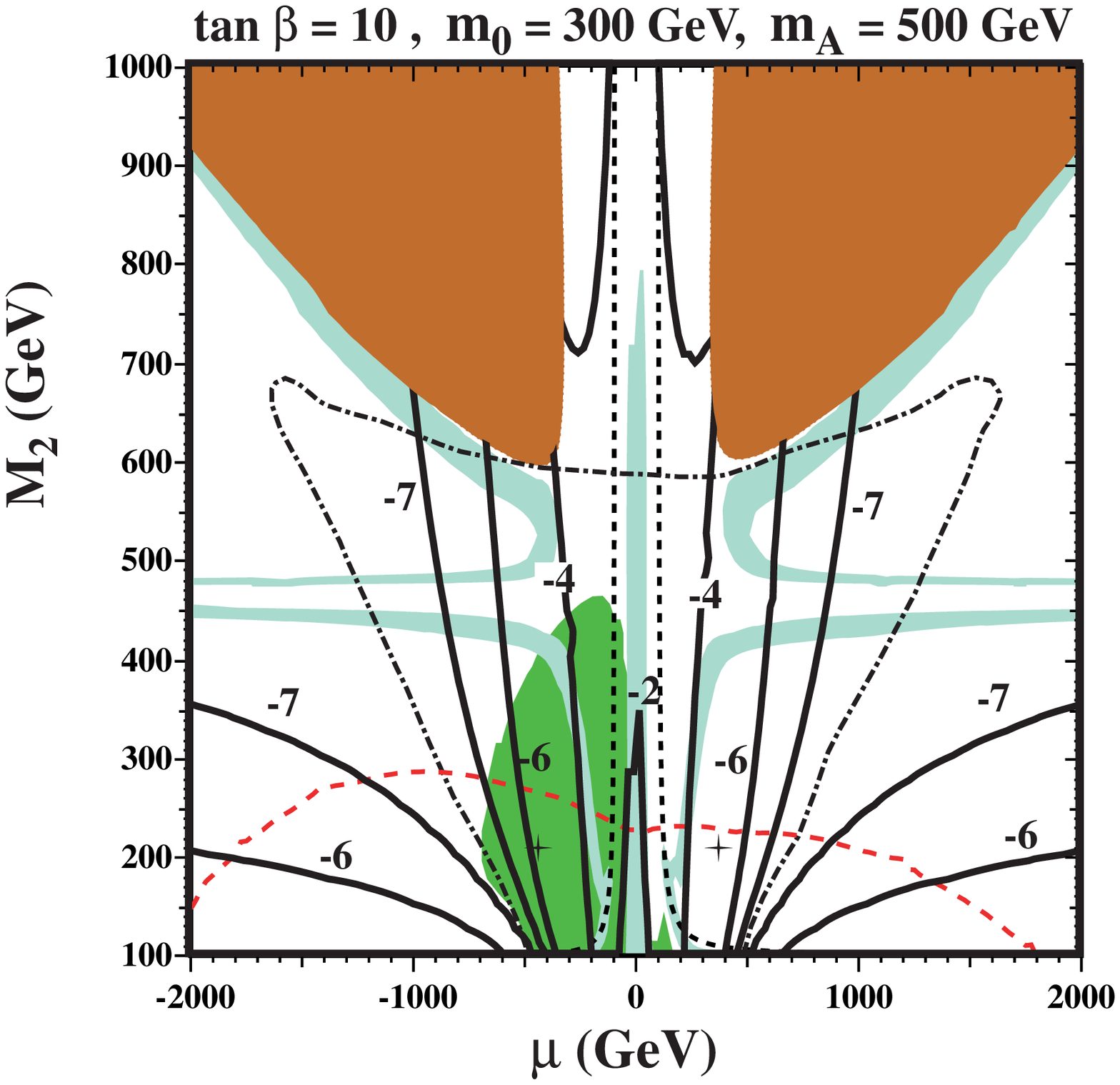,height=3.2in} \hfill
\end{minipage}
\caption{\label{fig:contourmuM2}
{\it 
Contours (a,b) of the spin-independent and (c,d) spin-dependent cross
sections (solid black lines) are superimposed on the NUHM $(\mu, M_2)$ 
planes for $\tan \beta = 10$, (a,c) $m_0 = 100$~GeV and
$\mA = 500$~GeV, (b,d) $m_0 = 300$~GeV and $\mA = 500$~GeV. The shadings and
line styles are the same as in Fig.~\ref{fig:contourmumA}, and the
crosses  denote the CMSSM points for these choices of $m_0$ and $m_A$.}}
\end{figure}

We display in Fig.~\ref{fig:contourmuM2} contours of (a,b) the
spin-independent and (c,d) the spin-dependent elastic scattering cross
sections, in the cases $\tan \beta = 10$, and (a,c) $m_0 = 100$~GeV and
$\mA = 500$~GeV, (b,d) $m_0 = 300$~GeV and $\mA = 500$~GeV. Because the
cross sections vary relatively rapidly, we have not included all the
decade cross-section contours in panels (c,d). 

We see again in this case the suppression in the spin-independent cross
section for $\mu \sim - 100$~GeV and $M_2 \gappeq 200$~GeV, apart from
which the cross sections decrease with increasing $|\mu|$, at least within
the GUT stability region. The spin-dependent cross section, on the other
hand, starts rising again at large $|\mu|$, reflecting the fact that one
has traversed a cancellation in the scattering matrix element of the type
described in the previous subsection. Once again, the cancellation is
found between the $10^{-7}$ pb contours just outside the GUT stability
curve.

For the values of parameters chosen in Fig.~\ref{fig:contourmuM2} (a,c),
we obtain a spin-independent cross section $\sigma_{SI}  = 2 \times
10^{-9}$ pb and a spin-dependent cross section $\sigma_{SD} = 3
\times 10^{-6}$ pb in the CMSSM (cf. the position of the crosses in the
figures).  In the NUHM, we find that the range of possible cross sections
for this case is $2.5 \times
10^{-11} {\rm pb} \la \sigma_{SI} \la  9 \times 10^{-8} {\rm pb}$
and $1.5 \times
10^{-9} {\rm pb} \la \sigma_{SD} \la   2 \times 10^{-3} {\rm pb}$.  For the
parameters in
Fig.~\ref{fig:contourmumA} (b,d), the CMSSM spin-independent
cross section is $\sigma_{SI}  = 3 \times 10^{-9}$ pb, while the
spin-dependent cross section is $\sigma_{SD} =  10^{-5}$ pb.  Note,
however, that this CMSSM point would be excluded due to a excessive value
for $\Omega h^2$ ( $\ga 1$). In the NUHM, this parameter 
choice is allowed
and gives the range  
$ 10^{-13} {\rm pb} \la \sigma_{SI} \la  2 \times 10^{-8} {\rm pb}$
and $9 \times 10^{-9} {\rm pb} \la \sigma_{SD} \la  2 \times 10^{-4} {\rm pb}$
for the elastic cross sections. 

\section{Allowed Ranges of Elastic Cross Sections}

Following our discussion in the previous Section of some important
features in a few examples of parameter planes in the NUHM, we now display
the ranges of elastic scattering cross sections permitted by various
theoretical and experimental constraints. We start with the specific NUHM
parameter planes discussed above, and then go on to generalize the
discussion. In each of the specific planes, we show the effect on the
allowed cross section when the phenomenological and cosmological
constraints are applied successively.  We start with the very simple
requirement that the LSP be a neutralino with $m_{\chi^\pm} \gappeq$ 103.5
GeV and $m_{\tilde e}\gappeq$ 99 GeV.  We then apply either the Higgs cut
or the $b \to s \gamma$ cut.  Our standard cut is defined to include these
two in addition to the appropriate value for $\Omega h^2$. Following the
standard cut we apply sequentially the GUT stability constraint and the
constraint due to $g_\mu - 2$.

\subsection{Specific Planes}

\begin{figure} 
\vspace*{-0.75in}
\begin{minipage}{6in}
\begin{center} 
\epsfig{file=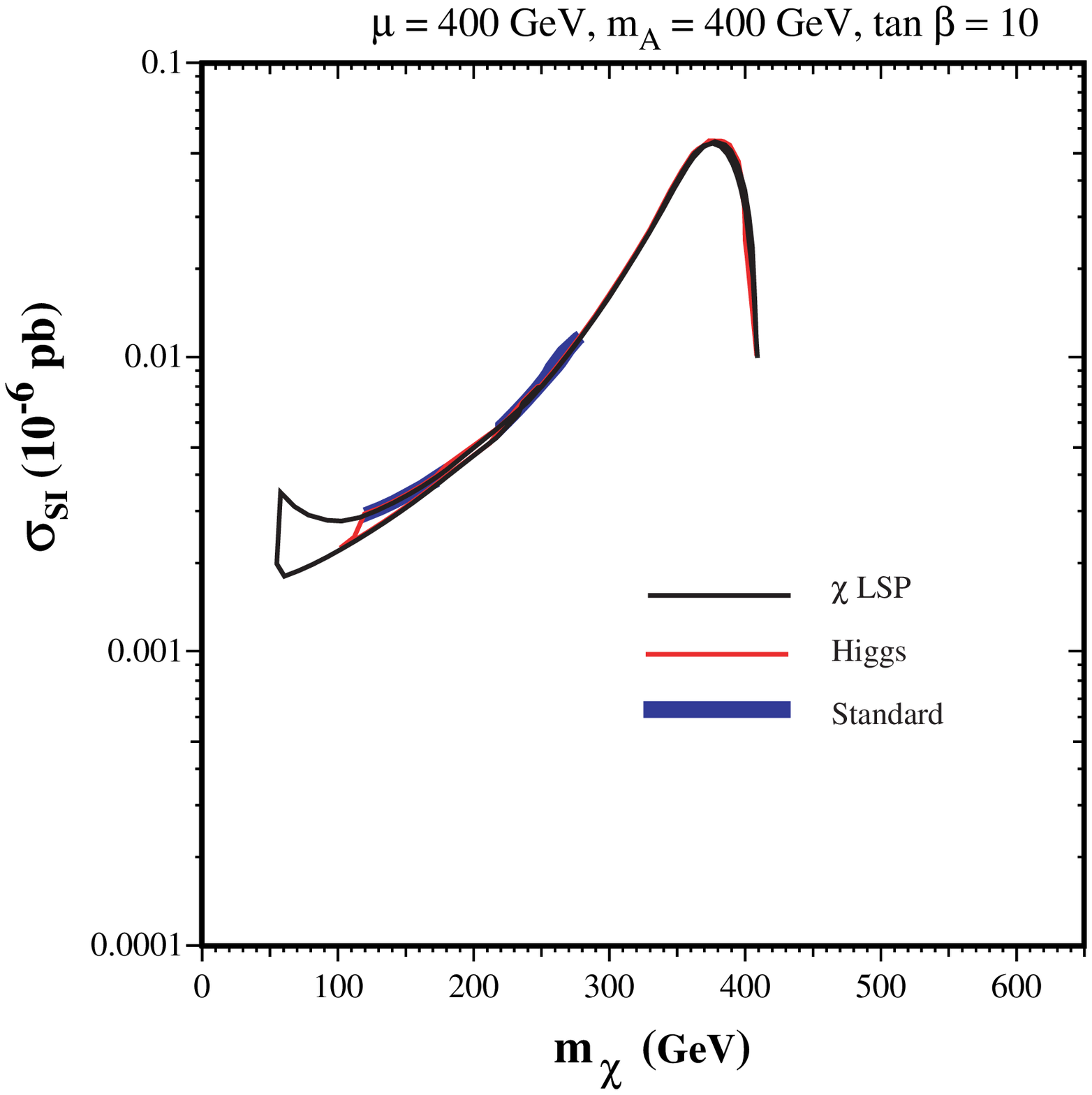,width=2.5in}
\hspace*{0.1in}
\epsfig{file=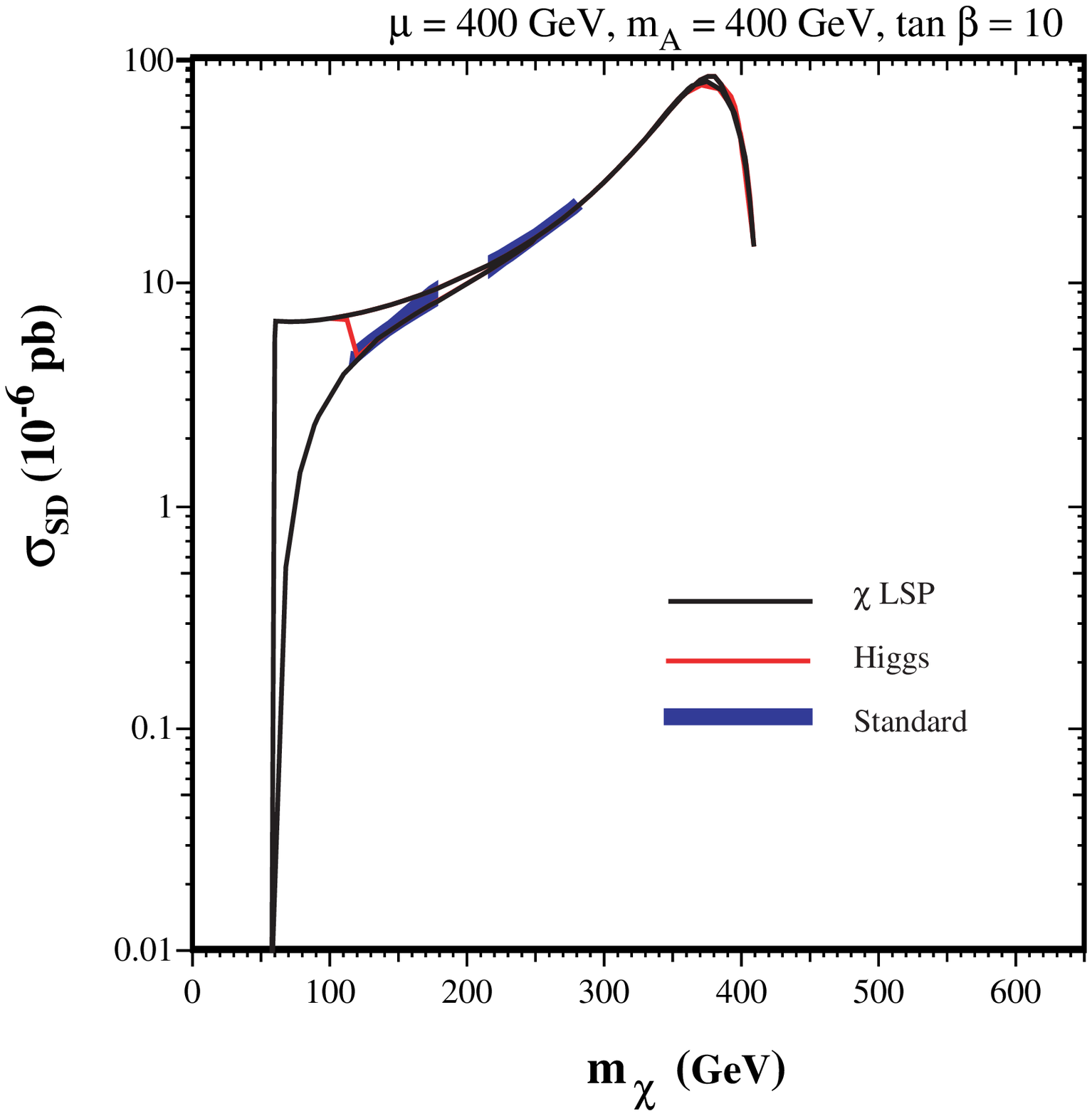,width=2.5in} \hfill
\end{center}
\end{minipage}
\begin{minipage}{6in}
\begin{center}
\epsfig{file=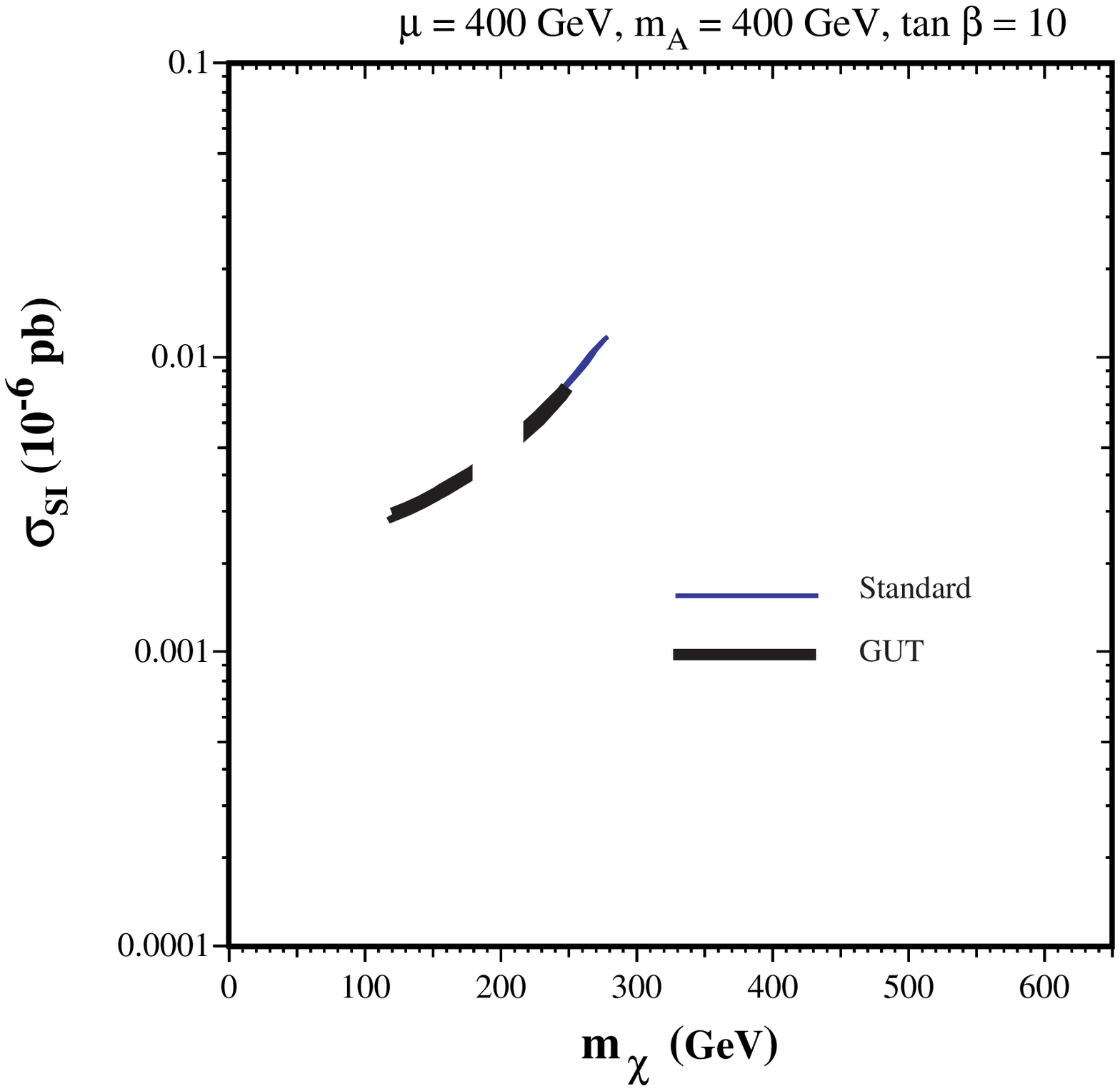,width=2.5in}
\hspace*{0.1in}
\epsfig{file=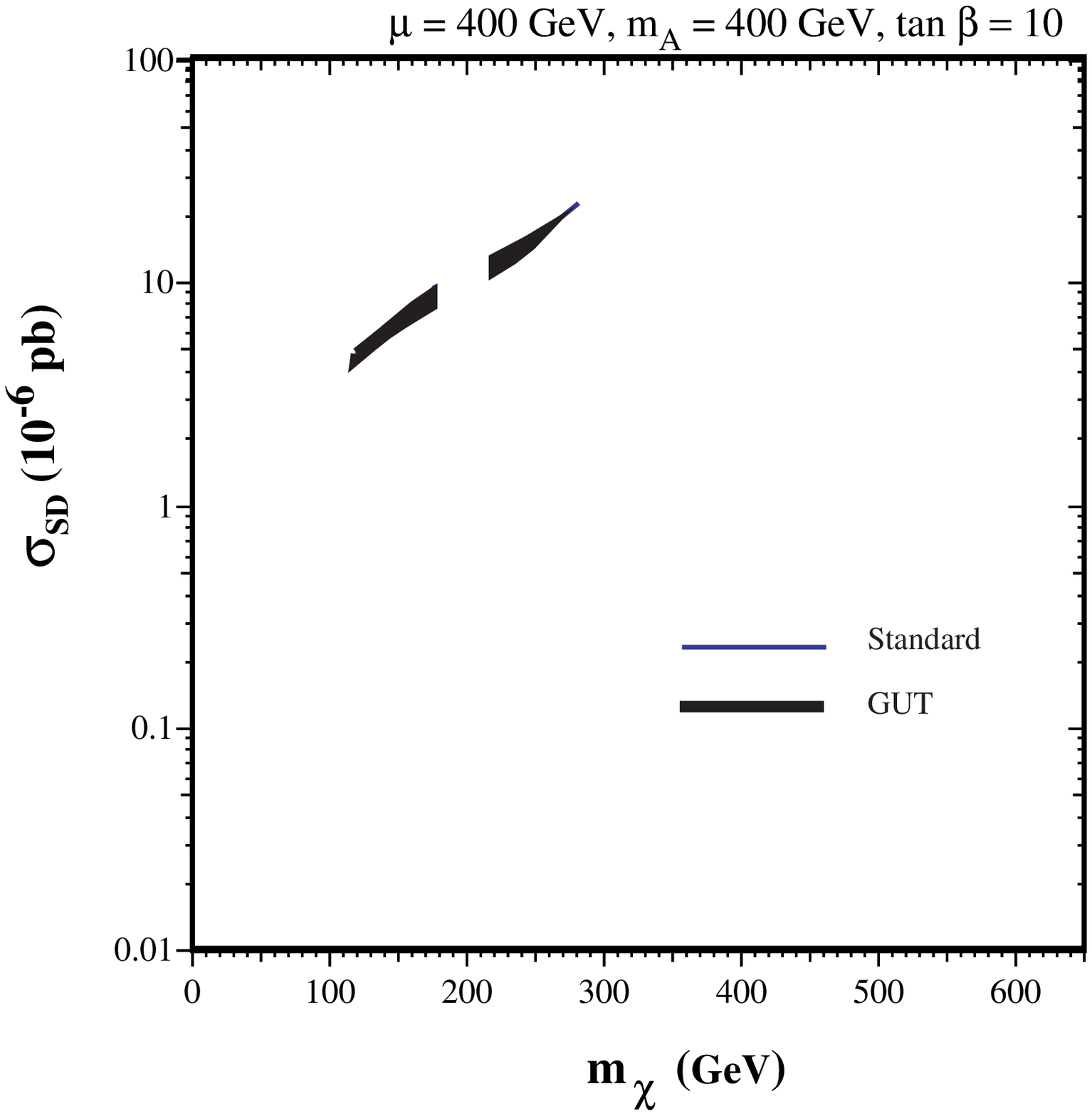,width=2.5in} \hfill
\end{center}
\end{minipage} 
\begin{minipage}{6in}
\begin{center} 
\epsfig{file=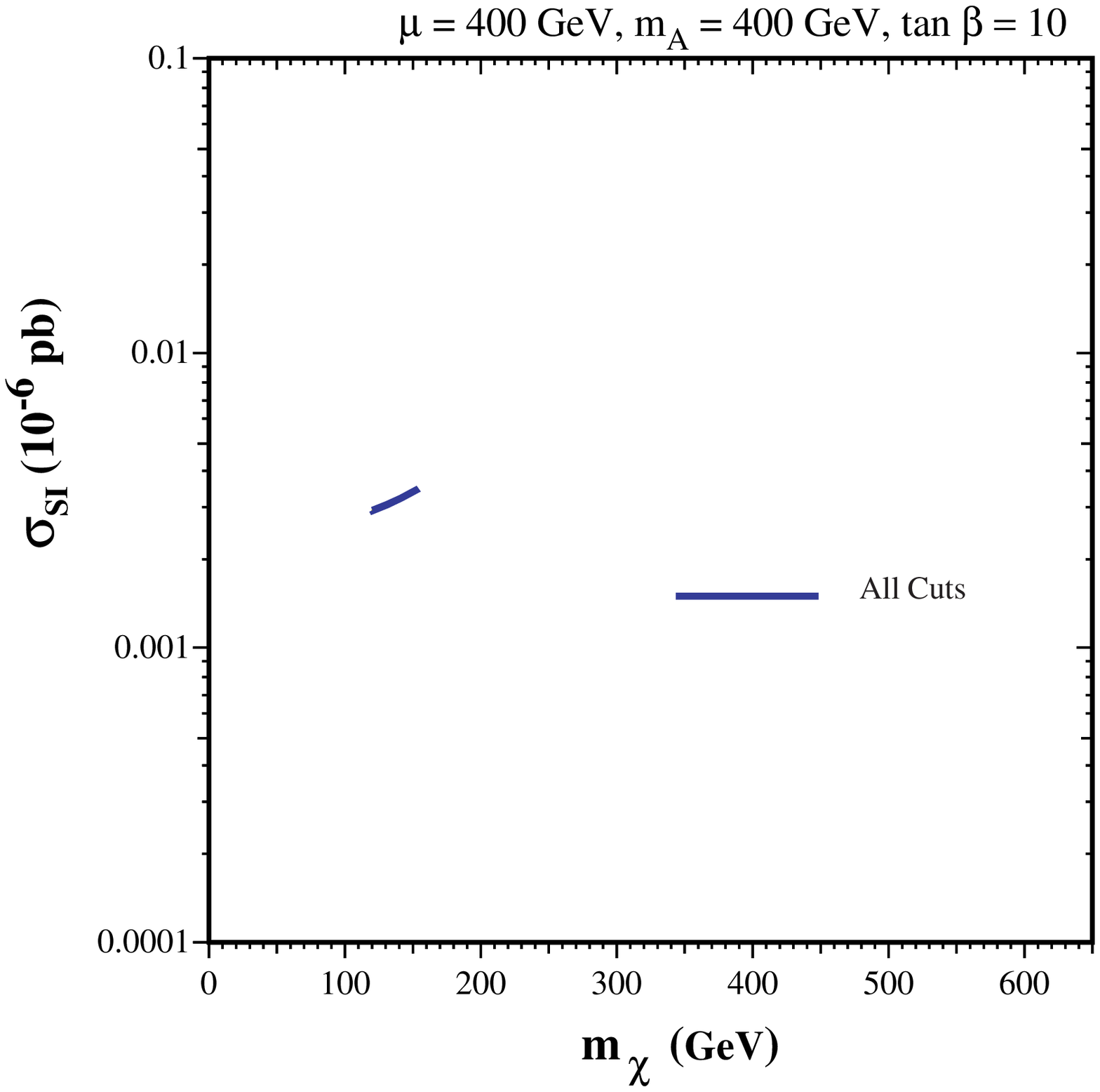,width=2.5in}
\hspace*{0.1in}
\epsfig{file=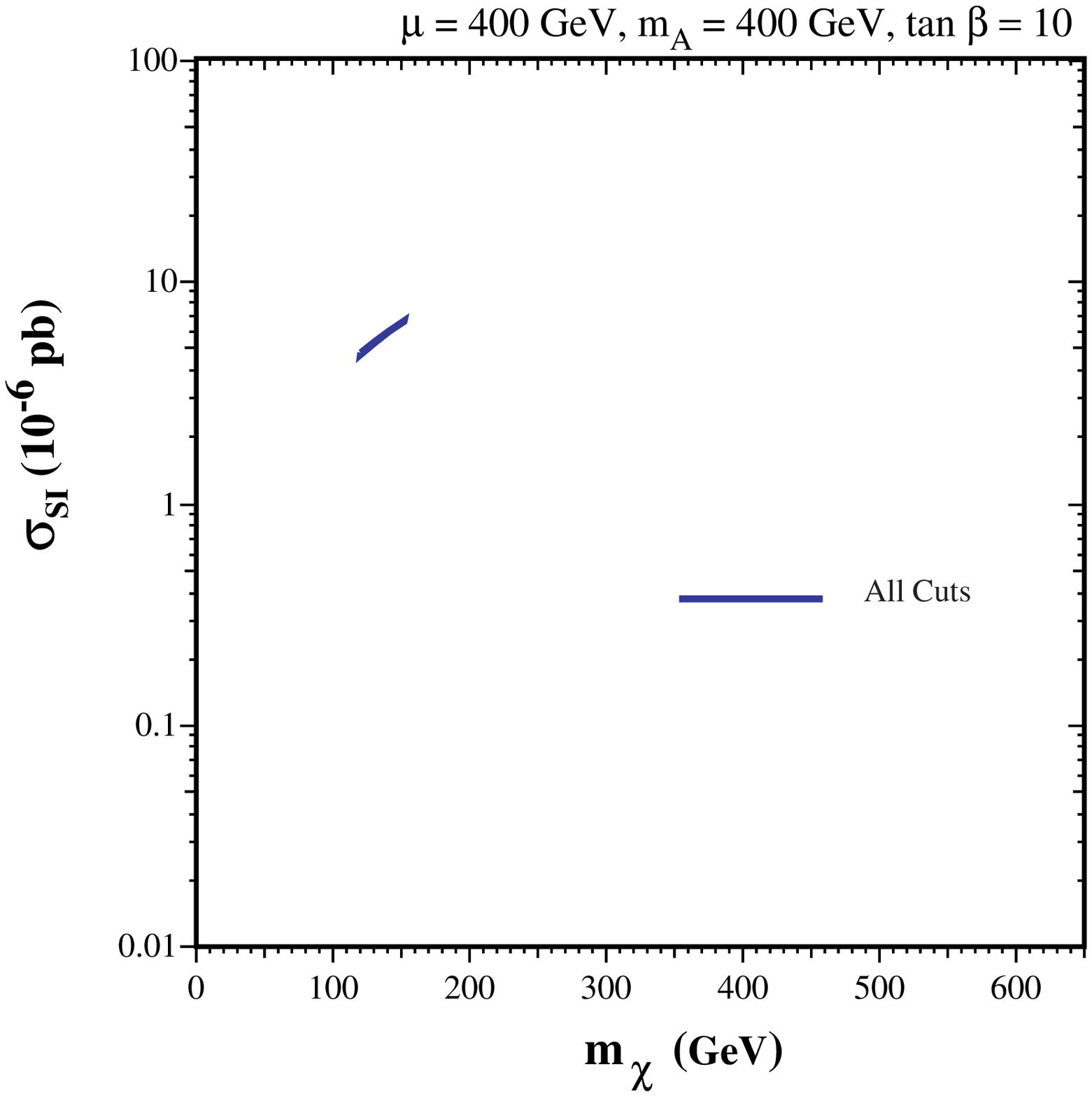,width=2.5in} \hfill
\end{center}   
\end{minipage}   
\caption{\label{fig:scatter2a}
{\it Cross sections allowed in a slice through the NUHM parameter space 
provided by the $(m_{1/2}, m_0)$ plane for $\tan \beta = 10$ and $\mu = 
m_A = 400$~GeV. Panels (a,b) are with our cuts on the LSP (dark 
lines), $m_h$ (lighter lines) and standard cut (shaded), panels (c,d) also
have the GUT stability 
constraint imposed, and panels (e,f) apply all cuts, including the 
possible $g_\mu - 2$ 
constraint. The right (left) panels show the spin-(in)dependent cross 
section, plotted against $m_\chi$.}} 
\end{figure}

We consider first the $(m_{1/2}, m_0)$ plane for $\tan \beta = 10$ and
$\mu = m_A = 400$~GeV that was displayed earlier in
Fig.~\ref{fig:contourmm}(a,c). The horizontal axes in the various panels
of Fig.~\ref{fig:scatter2a} correspond to the LSP mass $m_\chi$, and the
vertical axes show the ranges of (a,c,e) the spin-independent and (b,d,f)
the spin-dependent elastic cross sections.  The first row of panels (a,b)
shows the ranges allowed by our cuts on the LSP (dark lines),
$m_h$ (lighter lines) and standard cut (shaded), the second row (c,d)
displays the further impact of the GUT stability constraint, and the third
row (e,f) implements all the cuts, including that on $g_\mu - 2$.

We see in Fig.~\ref{fig:scatter2a}(a,b) that the scattering cross sections
rise monotonically with $m_\chi$, except close to the upper limit on
$m_\chi$, reflecting the increase with $m_{1/2}$ that was already
commented in connection with Fig.~\ref{fig:contourmm}.  We also note that
the ranges of cross sections allowed for any fixed value of $m_\chi$ is
very restricted, reflecting the fact the the contours of equal cross
section in Fig.~\ref{fig:contourmm}(a,c) are almost vertical in the
parameter range of interest. The break that appears in the middle of
Fig.~\ref{fig:scatter2a}(a,b), and is seen more clearly in (c,d), reflects
the range of $m_{1/2} \sim 500$~GeV where $\ohsq$ is suppressed below the
preferred cosmological range by rapid direct-channel $\chi \chi$
annihilation via the $H, A$ poles.

The most relevant effect of the extra GUT and $g_\mu - 2$ constraints is
to reduce the range of $m_{1/2}$ and hence $m_\chi$. As we see in
Fig.~\ref{fig:scatter2a}(c,d), the GUT stability constraint removes the
points in this NUHM parameter plane that have the largest elastic
scattering cross sections. Finally, as we see in
Fig.~\ref{fig:scatter2a}(e,f), the $g_\mu - 2$ constraint confines our
attention to points in the NUHM parameter below the rapid-annihilation
channel. In this particular case, there is a narrow preferred range of the
spin-independent cross section around $3 \times 10^{-9}$~pb, and the
preferred range of the spin-dependent cross section is around $6 \times
10^{-6}$~pb.

\begin{figure} 
\vspace*{-0.75in}
\begin{minipage}{6in}
\begin{center} 
\epsfig{file=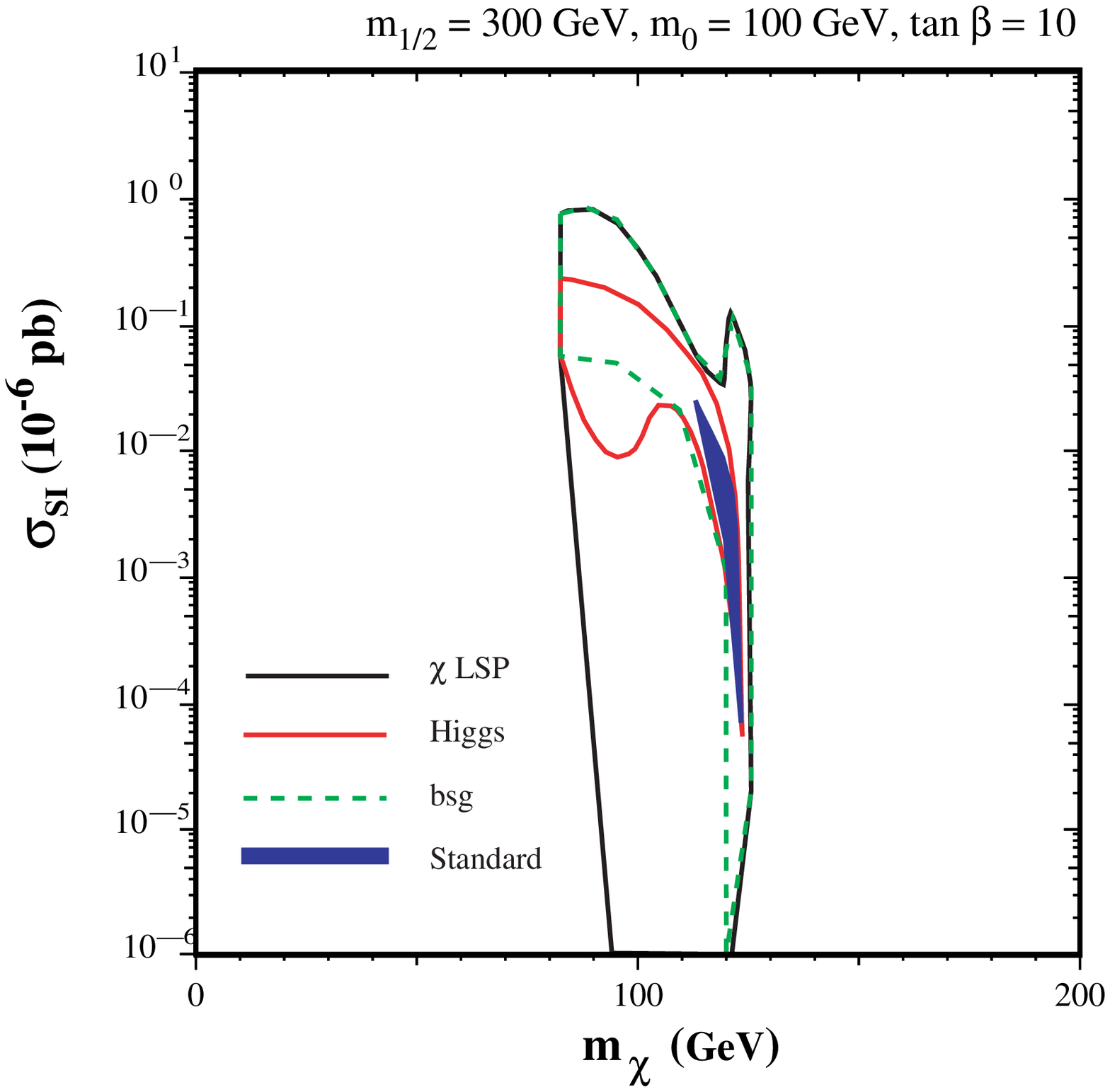,width=2.5in}
\hspace*{0.1in}
\epsfig{file=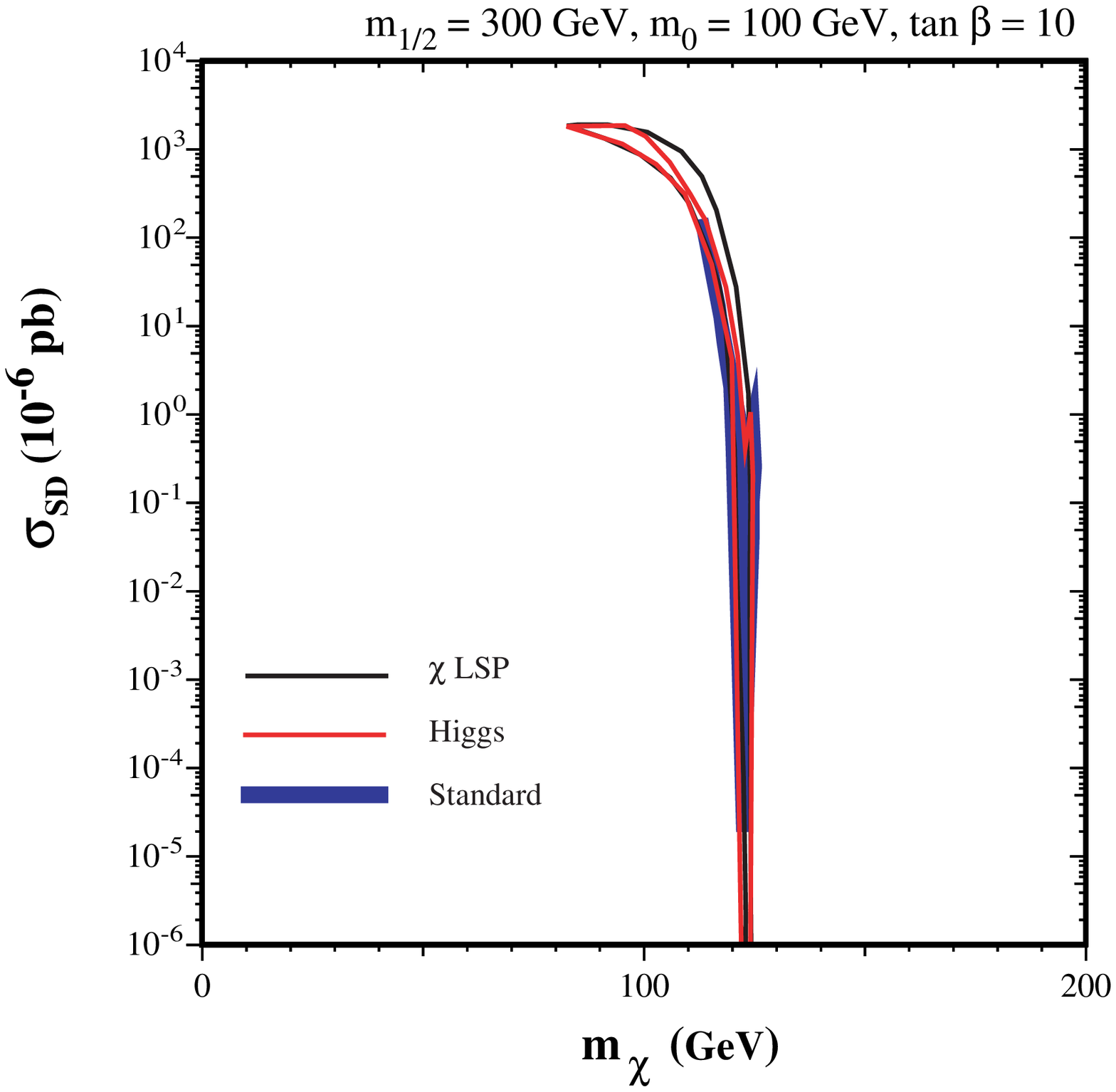,width=2.5in} \hfill
\end{center}
\end{minipage}
\begin{minipage}{6in}
\begin{center}
\epsfig{file=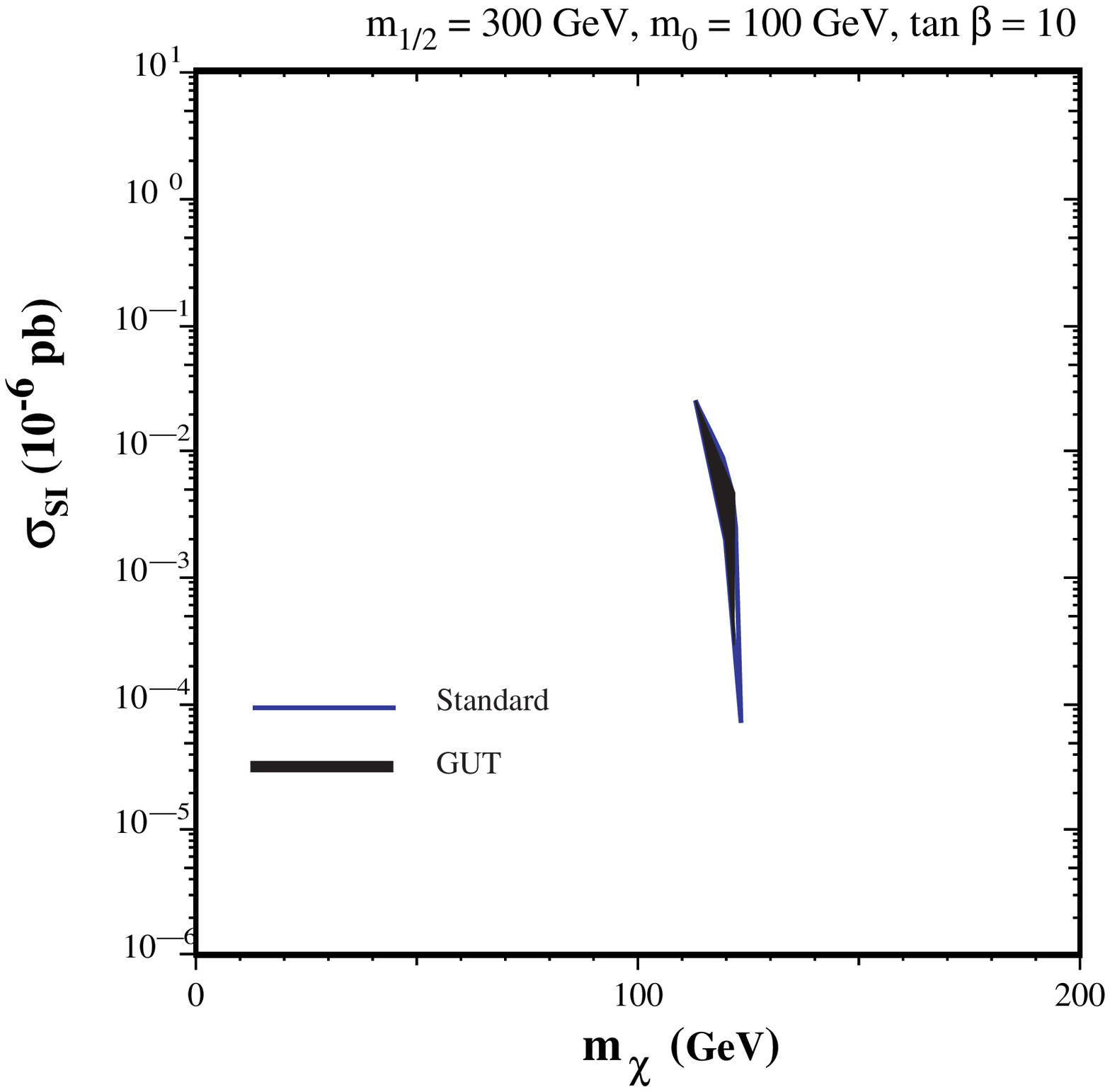,width=2.5in}
\hspace*{0.1in}
\epsfig{file=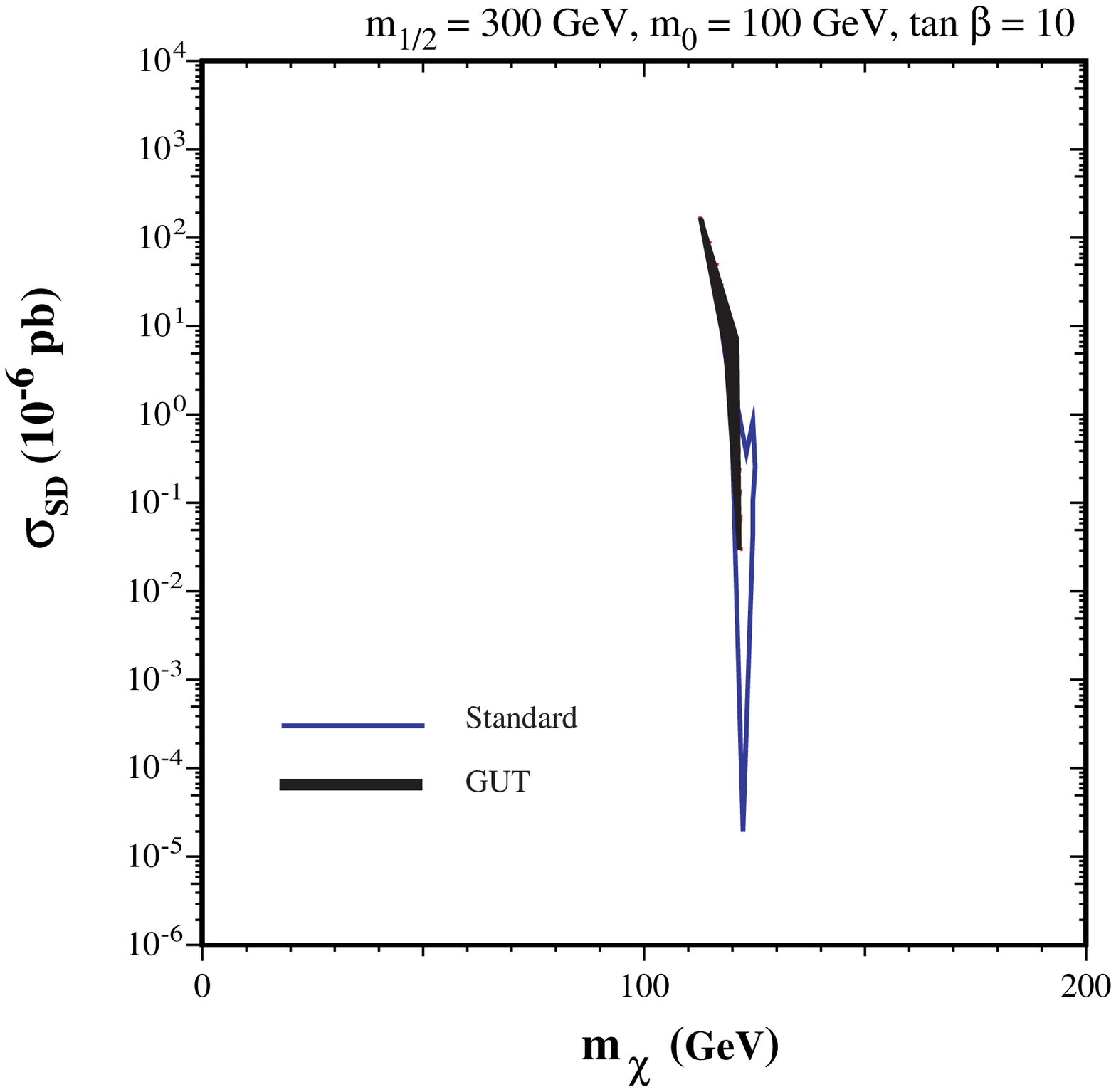,width=2.5in} \hfill
\end{center}
\end{minipage} 
\begin{minipage}{6in}
\begin{center} 
\epsfig{file=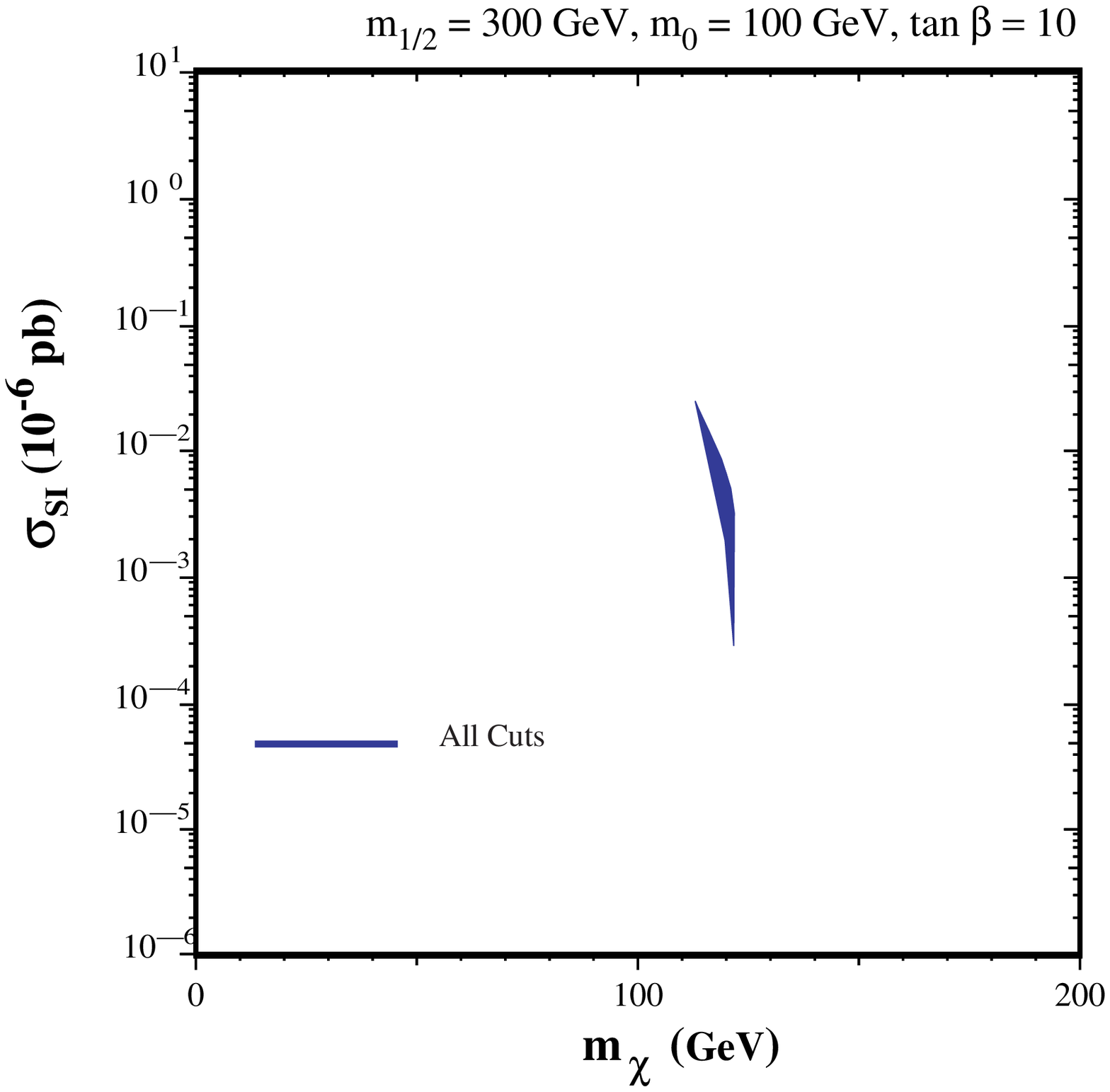,width=2.5in}
\hspace*{0.1in}
\epsfig{file=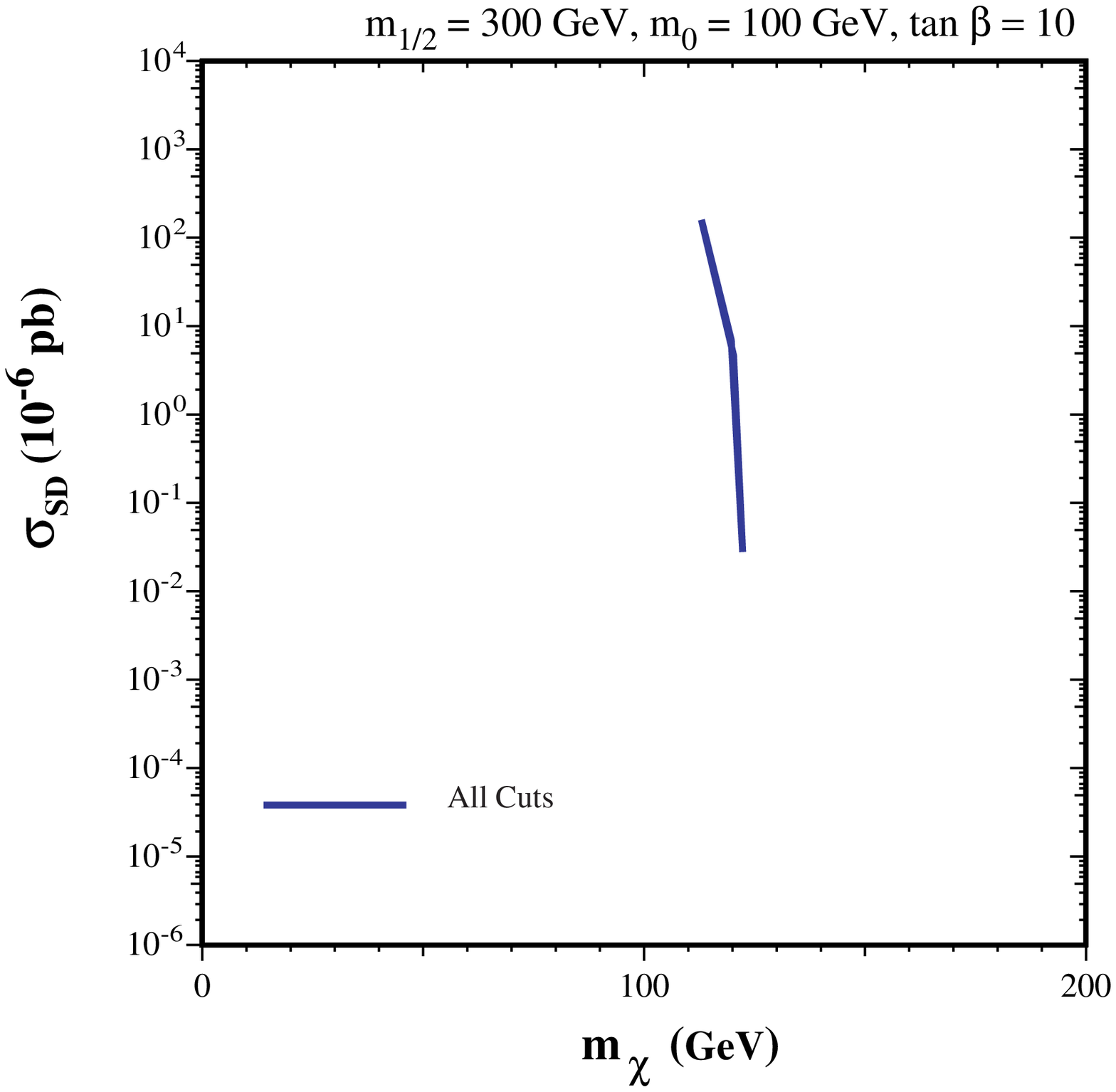,width=2.5in} \hfill
\end{center}   
\end{minipage}   
\caption{\label{fig:scatter4a}
{\it Cross sections allowed in a slice through the NUHM parameter space 
provided by the $(\mu, m_A)$ plane for $\tan \beta = 10$ and $m_{1/2} = 
300$~GeV, $m_0 = 100$~GeV. The selections of points in the different 
panels are the same as in Fig.~\ref{fig:scatter2a}, as are their 
indications. In addition, panel (a) also shows the $b \to s \gamma$ cut (dashed
lines).}}
\end{figure}

Our second example is the $(\mu, m_A)$ plane for $\tan \beta = 10, m_{1/2}
= 300$~GeV and $m_0 = 100$~GeV, displayed previously in panels (a,c) of
Fig.~\ref{fig:contourmumA}. As seen in Fig.~\ref{fig:scatter4a}(a,b), when
one imposes the standard $\ohsq, m_h$ and $b \to s \gamma$ constraints,
the cross sections generally decrease with $m_\chi$. However, in the
spin-dependent case the cross section reaches a locus of near zeroes,
after which it rises again. These cancellations are avoided in this case
when the Higgs and $b \to s \gamma $ constraints are applied. As then seen
in Fig.~\ref{fig:scatter4a}(c,d), the GUT stability constraint, which
removes portions of this NUHM parameter plane at large
$|\mu|$ and small
$m_A$, strengthens in this case the lower bounds on the cross sections.
The
effect of the $g_\mu - 2$ constraint is less marked in this case, as seen
in Fig.~\ref{fig:scatter4a}(e,f). The final allowed ranges of the cross
section are considerably wider than in the previous example: $\sim
10^{-9}$ to $\sim 10^{-8}$~pb in the spin-independent case and $\sim
10^{-7}$ to $\sim 10^{-4}$~pb in the spin-dependent case.

\begin{figure} 
\vspace*{-0.75in}
\begin{minipage}{6in}
\begin{center} 
\epsfig{file=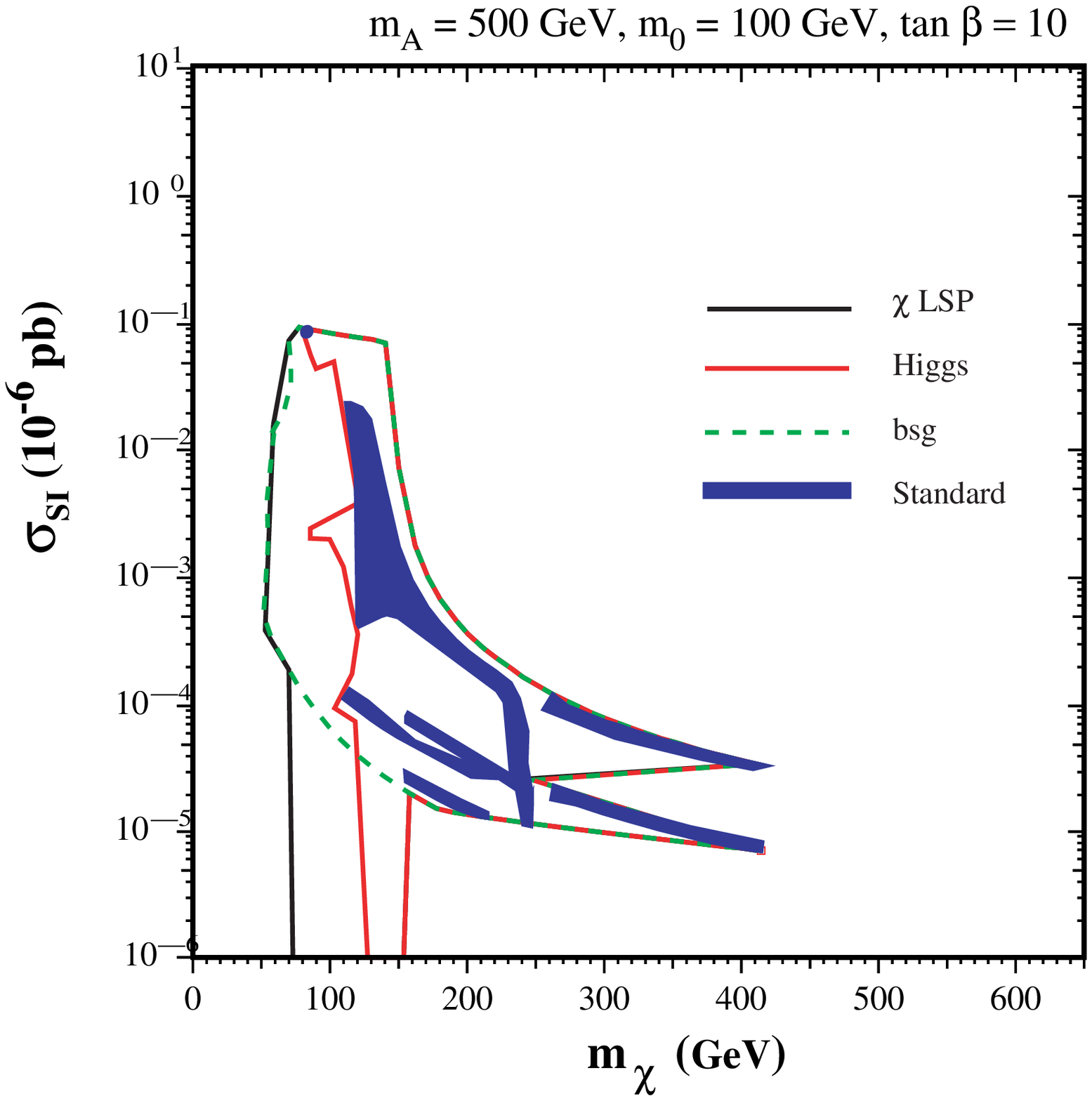,width=2.5in}
\hspace*{0.1in}
\epsfig{file=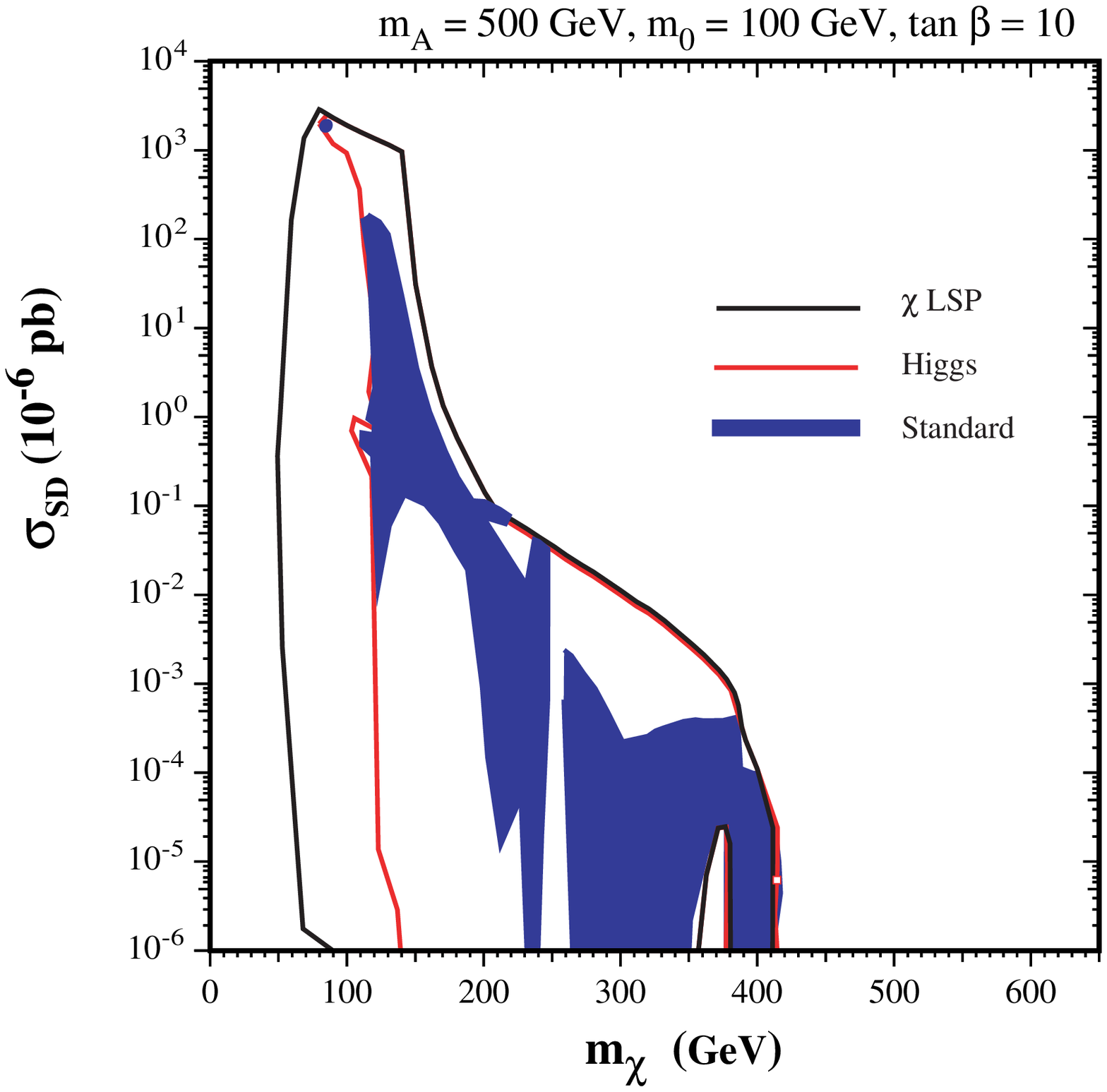,width=2.5in} \hfill
\end{center}
\end{minipage}
\begin{minipage}{6in}
\begin{center}
\epsfig{file=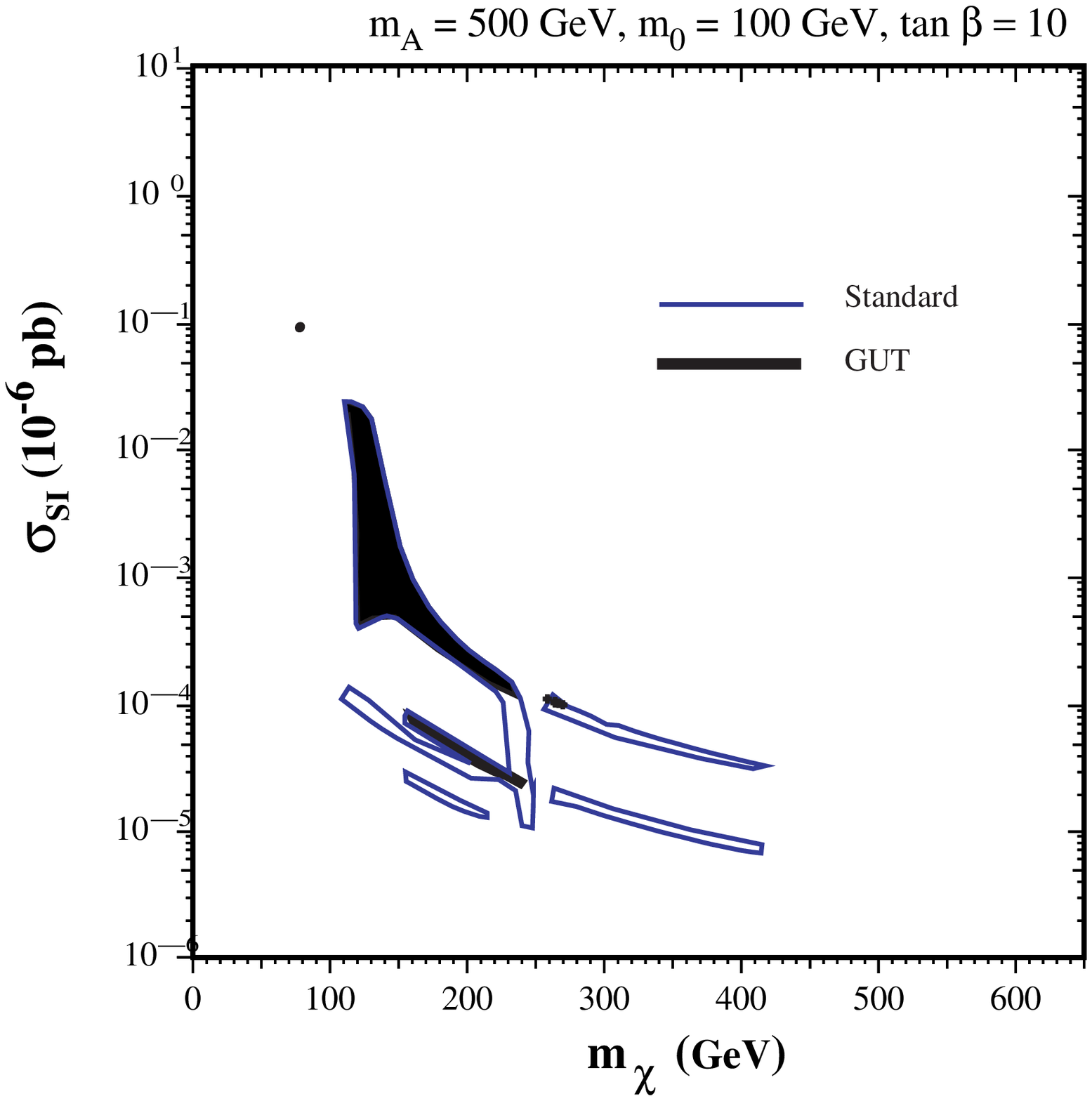,width=2.5in}
\hspace*{0.1in}
\epsfig{file=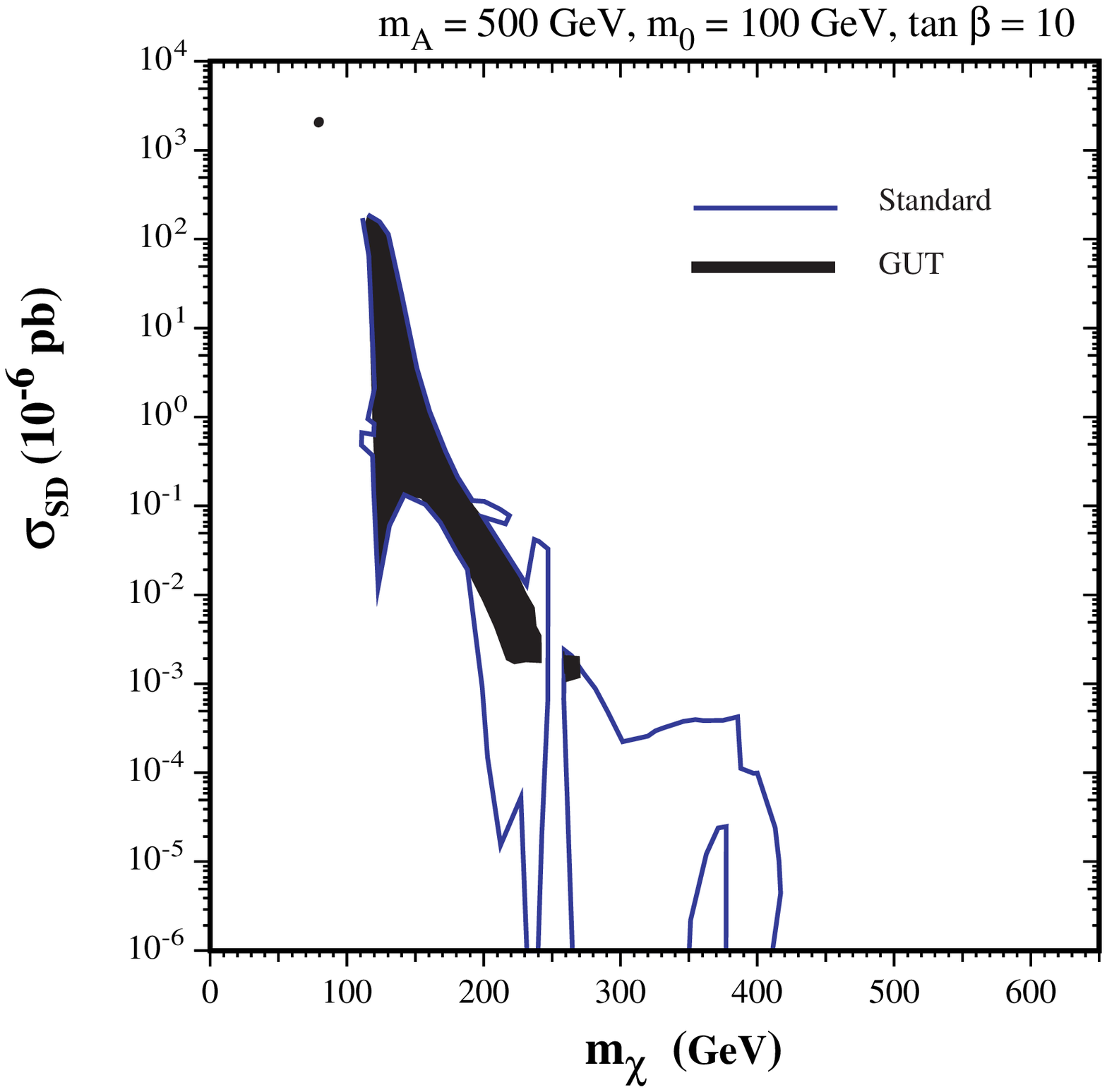,width=2.5in} \hfill
\end{center}
\end{minipage} 
\begin{minipage}{6in}
\begin{center} 
\epsfig{file=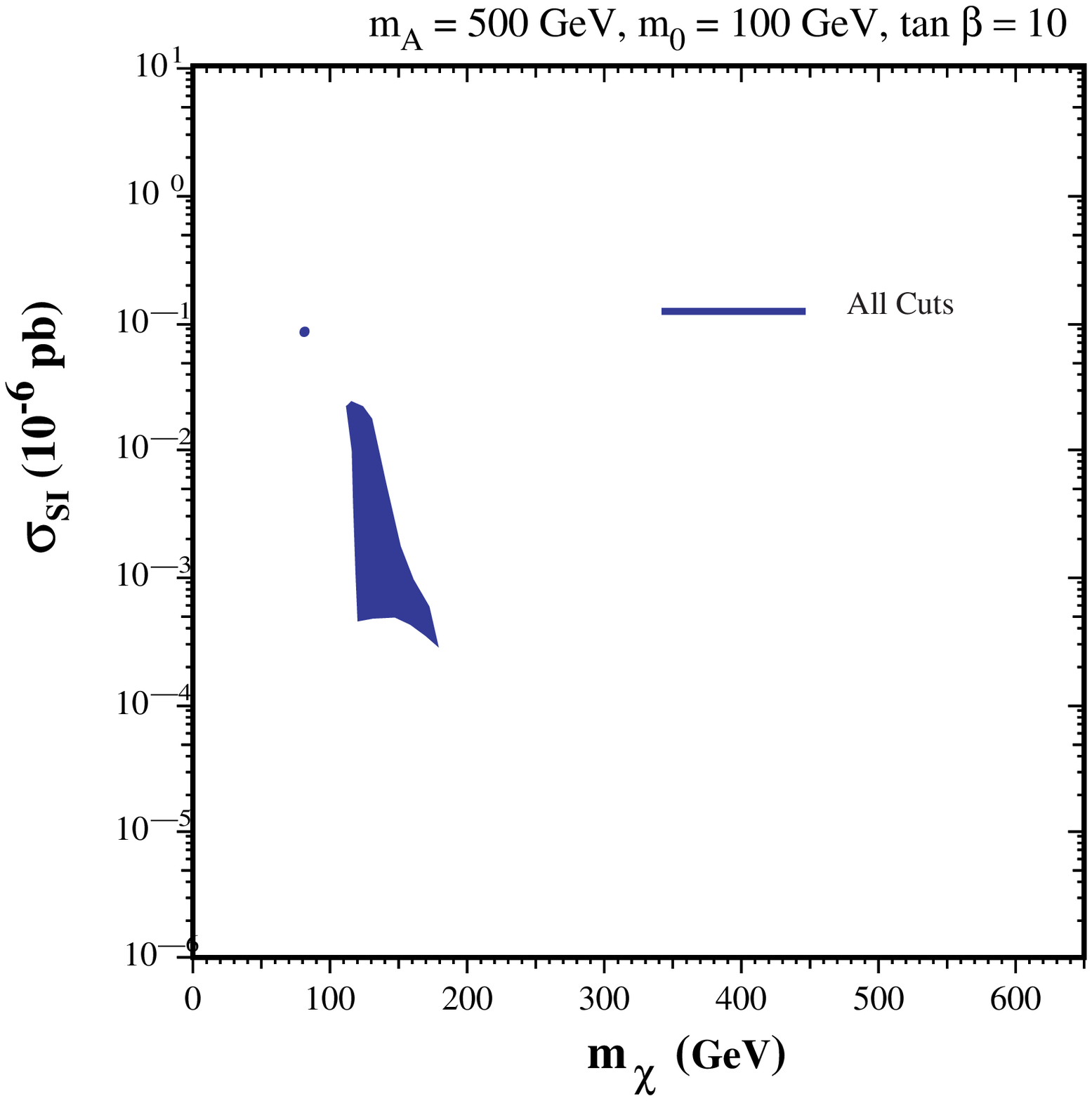,width=2.5in}
\hspace*{0.1in}
\epsfig{file=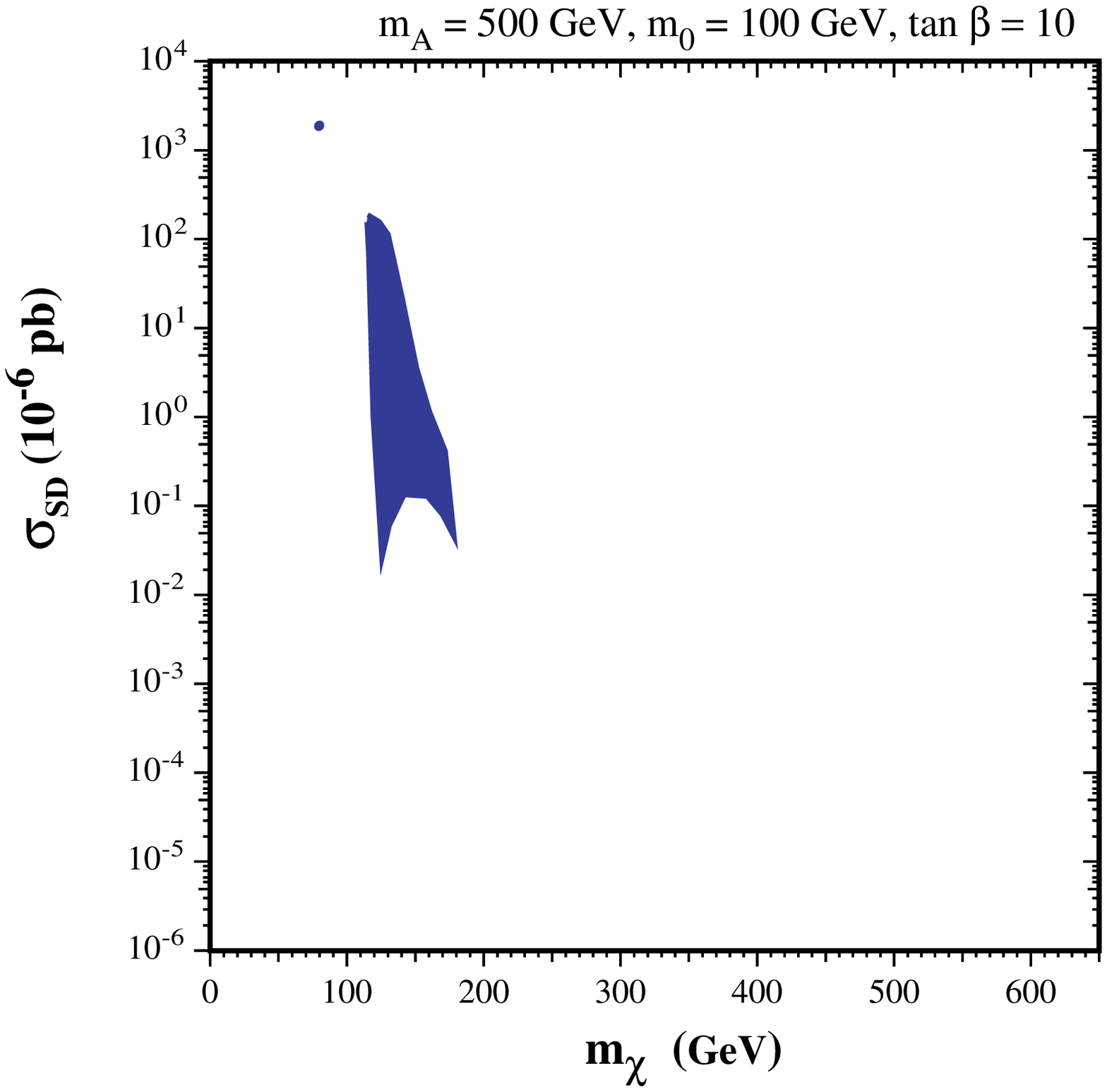,width=2.5in} \hfill
\end{center}   
\end{minipage}   
\caption{\label{fig:scatter8a}
{\it  Cross sections allowed in a slice through the NUHM parameter space
provided by the $(\mu, M_2)$ plane for $\tan \beta = 10$ and $m_0 =   
100$~GeV, $m_A = 500$~GeV. The selections of points in the different
panels are the same as in Fig.~\ref{fig:scatter4a}, as are their 
indications. }}
\end{figure}

Our final example is the $(\mu, M_2)$ plane for $\tan \beta = 10, m_0 =
100$~GeV and $m_A = 500$~GeV, shown in Fig.~\ref{fig:scatter8a}. In this
case, the standard cuts allow a particularly wide range of cross
sections, varying by infinite (over 3) orders of magnitude for the
spin-(in)dependent case.  The different regions allowed by the standard
cuts reflect the different branches of parameter space in
Fig.~\ref{fig:contourmuM2}. We note that some of these branches are due
to the differences between positive and negative $\mu$.  We also note
that some of the boundaries are due to our imposed cutoff of $|\mu| \le
2$ TeV.  In particular, had we allowed for larger values of $|\mu|$, we
would have found larger neutralino masses, and the lower bounds for 
both the Higgs and $b \to s \gamma$ cuts (which differ for
positive and negative $\mu$) would also have been lowered. The general 
tendency
of the standard cuts is to decrease the cross sections with increasing
$m_\chi$, though with considerable variation. The GUT stability
constraint in this case removes a region at large $M_2$ and hence
$m_\chi$, that removes the points with the lowest cross sections. This
effect is particularly marked for the spin-dependent case, where the
range is now reduced to (a mere) 5 orders of magnitude. The $g_\mu - 2$
constraint further raises the lower bounds on the cross sections, so that
they vary through just over 3 (under 2) orders of magnitude in the
spin-(in)dependent case.

Notice that there is an isolated point at $m_\chi \sim$ 80 GeV. This
corresponds to a very narrow region just to the right of the chargino 
mass-bound
line in Fig.~\ref{fig:contourmuM2}(a,c) at $\mu \sim$ 120 GeV, $M_2 \sim$ 240
GeV, which is not visible because of the plotting resolution. What happens 
is that as $\mu$
decreases,
$m_\chi$ falls below the $m_h$ threshold, and hence the annihilation cross
section decreases leading to an acceptable value of $\ohsq$. However, as $\mu$
decreases further, neutralino-chargino coannihilation becomes stronger,
suppressing $\ohsq$ again to be less than 0.1.  

This brief survey shows the importance of implementing correctly the GUT
stability constraint, which may (in different cases) bound the cross
sections either above or below. As many authors have previously pointed
out in the CMSSM case, the $g_\mu - 2$ constraint is also potentially
important. In certain cases, it can also strengthen significantly the
lower limits on the NUHM cross sections.

\subsection{General Analysis}

\begin{figure}
\vspace*{-0.75in}
\hspace*{.70in}
\begin{minipage}{8in}
\epsfig{file=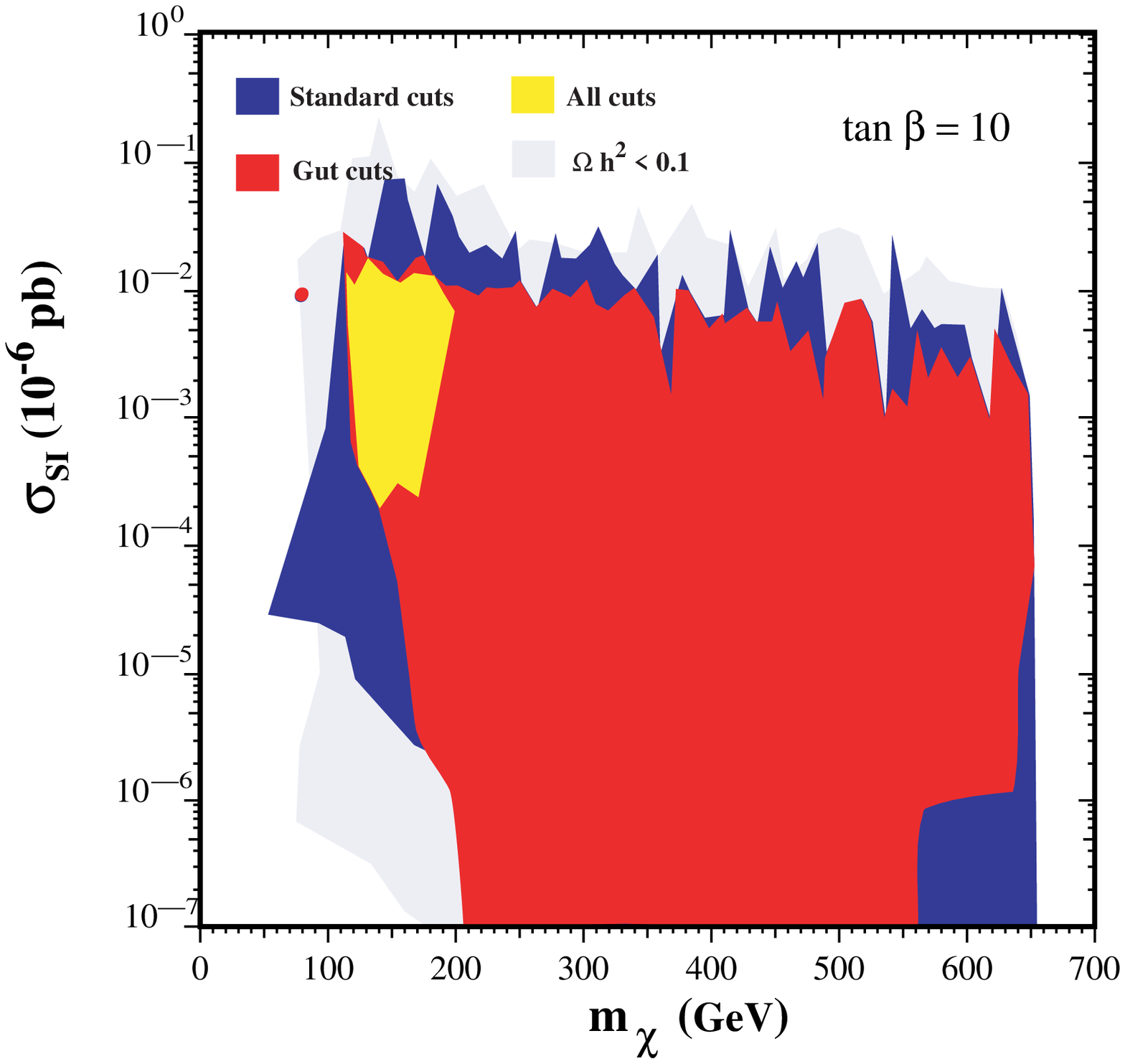,height=4in}
\epsfig{file=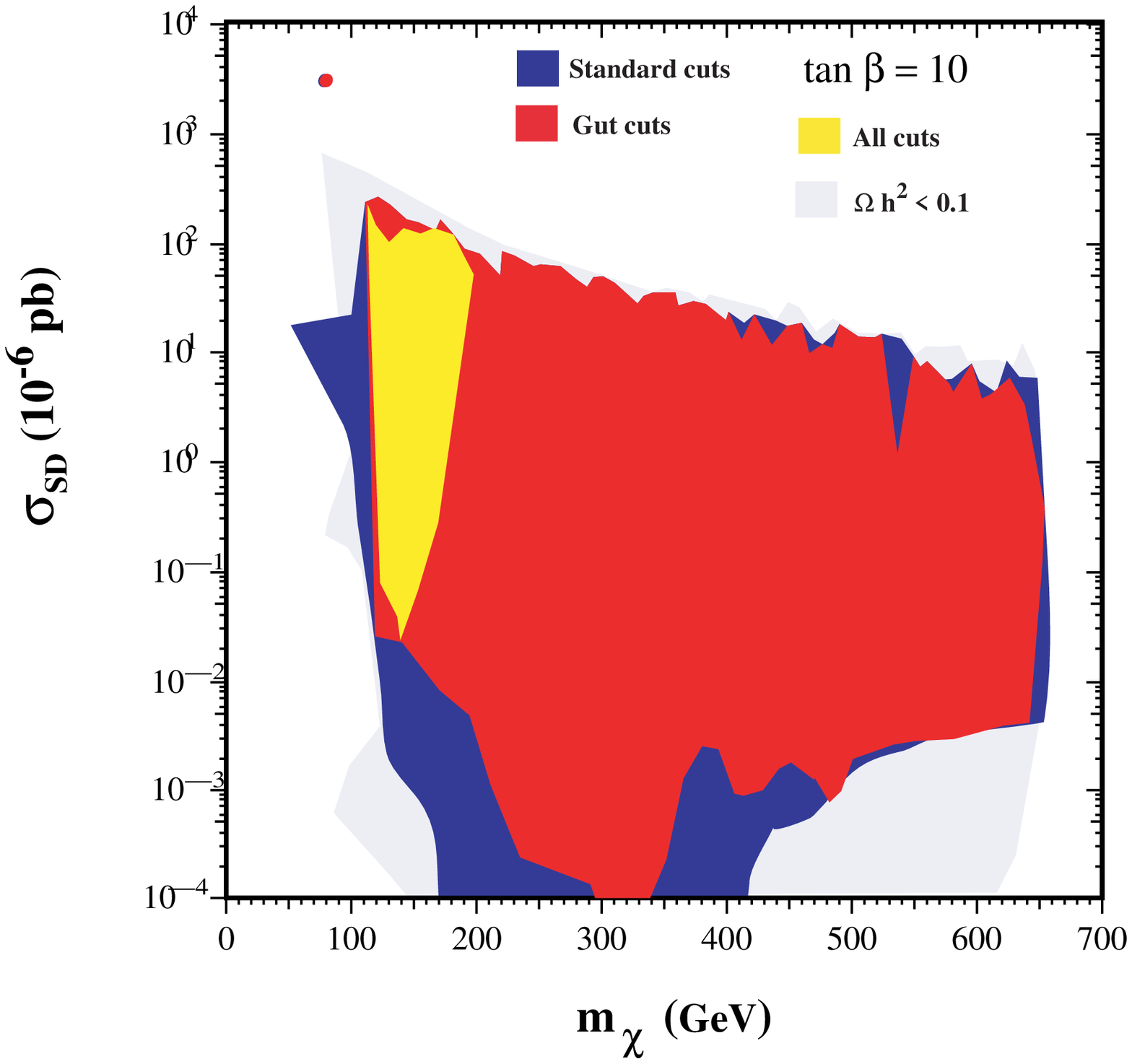,height=4in} \hfill
\end{minipage}
\caption{\label{fig:Andy}
{\it 
Ranges of (a) the spin-independent and (b) the spin-dependent 
cross sections for $\tan \beta = 10$. The ranges allowed by the standard 
cuts on $\ohsq$, $m_h$ and $b \to s \gamma$ have dark shading, those still 
allowed by the GUT stability cut have medium shading, and those still 
allowed after applying all the cuts including $g_\mu - 2$ have light 
shading.   The pale shaded region
corresponds to the extra area of points with low relic 
densities, whose cross sections have been rescaled appropriately.}} 
\end{figure}

Equipped with the above information about some specific examples, we now
make a general analysis of the possible values of the elastic scattering
cross sections. In the first place, we concentrate on the case
$\tan \beta = 10$, but relaxing the previous restricted choices of 
other parameters that we took as examples. To produce the plots, we generate
random points (about 30000 points for each plot) within 
the following ranges:
\begin{eqnarray}
100 \; {\rm GeV} \leq m_{1/2} \leq 1500 \; {\rm GeV}, \nnl
0 \leq m_0 \leq 1000 \; {\rm GeV}, \nnl
-2000 \; {\rm GeV} \leq \mu \leq 2000 \; {\rm GeV}, \nnl
90 \; {\rm GeV} \leq m_A \leq 1500 \; {\rm GeV}.
\label{ranges}
\end{eqnarray}
We first impose the same standard experimental and phenomenological 
constraints discussed earlier, namely: a consistent electroweak vacuum,
$0.1 < \ohsq < 0.3$, $m_h > 114$~GeV and the 
$b \to s \gamma$ constraint. In a previous paper on the CMSSM~\cite{EFlO2}, we 
rescaled the elastic scattering cross sections for models that predicted 
$\ohsq < 0.1$ by the factor $\ohsq / 0.1$, so as to account for the fact 
that the neutralino could constitute at most this fraction of the galactic
halo. In the results that follow, we show how this rescaling affects
the upper limits on the cross sections. Next we apply the GUT stability
cut, and  finally we show the implications of imposing the cut on $g_\mu
- 2$. The resulting ranges of the elastic scattering cross sections 
for $\tan \beta = 10$ are displayed in Fig.~\ref{fig:Andy}.

We note that the spin-independent cross section shown in
Fig.~\ref{fig:Andy}(a) may be as large as a few $\times 10^{-8}$~pb,
decreasing only slightly as $m_\chi$ increases, whilst values lower
than $10^{-13}$~pb cannot be excluded. The raggedness of
the upper limit on the cross section reflects the fact that our sampling
produced very few points with such large cross sections: values in between
the crags cannot be excluded, but must be very rare. 
Between the crags there are valleys, below which the density of points is
significantly larger.  The lowest values of the cross section occur for
$\mu < 0$, where cancellations are possible in the spin-independent
scattering matrix element, as discussed in a previous paper~\cite{EFlO1}.
The GUT stability constraints exclude some low cross-section values at
both small and large
$m_\chi$, but do not provide an overall lower bound. It does however,
lower the upper bound by as much as a factor of about 5. Negative $\mu$,
and hence very low cross-section values, would be excluded by the
putative $g_\mu -2$ constraint, as shown in Fig.~\ref{fig:Andy}(a). For
$\mu > 0$, we find (not shown) spin-independent cross sections only above
$10^{-11}$~pb.

Fig.~\ref{fig:Andy}(a) also displays the region (pale shaded) which
survives all standard cuts, except that $\Omega h^2 < 0.1$.
For points in this region, we have rescaled the cross section by a factor
of $\Omega h^2/0.1$, to allow for the fact that the LSP could not in this 
case make up all the cold dark matter in the Universe, and hence 
{\it a fortiori} in the galactic halo. As one can see, many of the 
spaces between the crags are now
filled in with such points, but very few give significantly larger cross
sections. 

In the case of the spin-dependent cross section shown in
Fig.~\ref{fig:Andy}(b), the upper limit is better defined, and decreases
monotonically from $\sim 3 \times 10^{-3}$~pb for $m_\chi \sim 80$~GeV to
$\sim 5 \times 10^{-6}$~pb for $m_\chi \sim 650$~GeV. Cross sections lower
than $\sim 10^{-10}$~pb are possible for either sign of $\mu$, even after
imposing the GUT stability cuts. In this case, the points with rescaled
cross sections enhance the cross section by a factor of about 3 at low
neutralino masses.

The isolated point in both panels at $m_\chi \sim$ 80 GeV now corresponds
to a narrow region around $\mu \sim -110$ GeV and large $m_A > 1000$ GeV,
which is between the $m_h = 114$ GeV and $m_{\chi^\pm} = 103.5$ GeV lines.
Its existence is very sensitive to the implementation of the Higgs mass 
bound.

We note that the ranges allowed by $g_\mu - 2$ are relatively restricted.  
For a start, we find that $110$~GeV $\lappeq m_\chi \lappeq 200$~GeV.  
Moreover, even within this range, very low cross-section values are
excluded.   Overall, we find ranges between $\sim 2 \times 10^{-8}$~pb
and $\sim 2
\times 10^{-10}$~pb for the spin-independent cross section, and between
$\sim 2 \times 10^{-4}$~pb and $\sim 2 \times 10^{-8}$~pb for the
spin-independent cross section.

\begin{figure}
\vspace*{-0.75in}
\hspace*{.70in}
\begin{minipage}{8in}
\epsfig{file=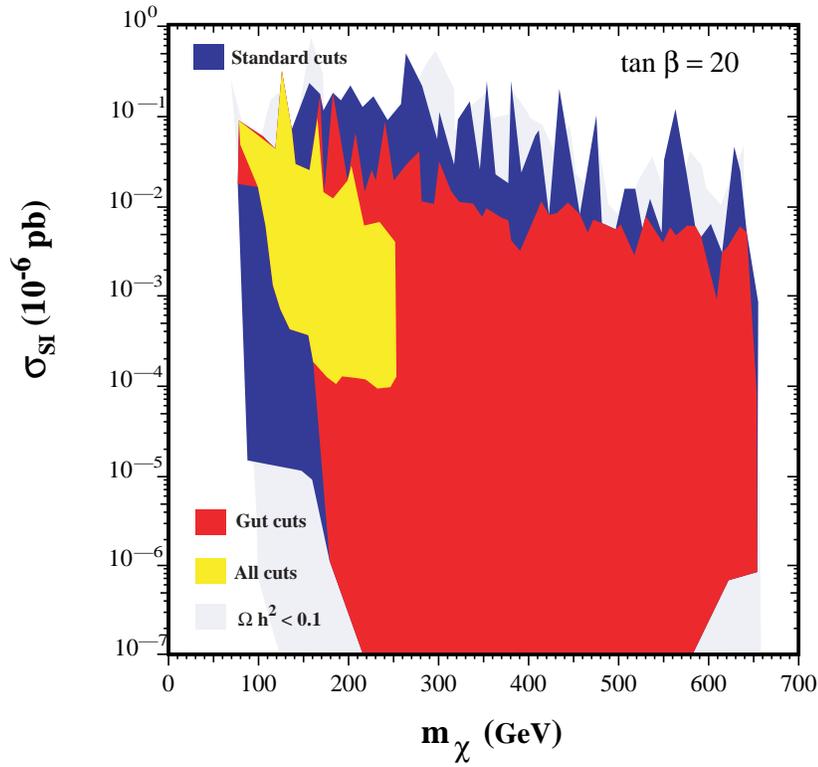,height=4in}
\epsfig{file=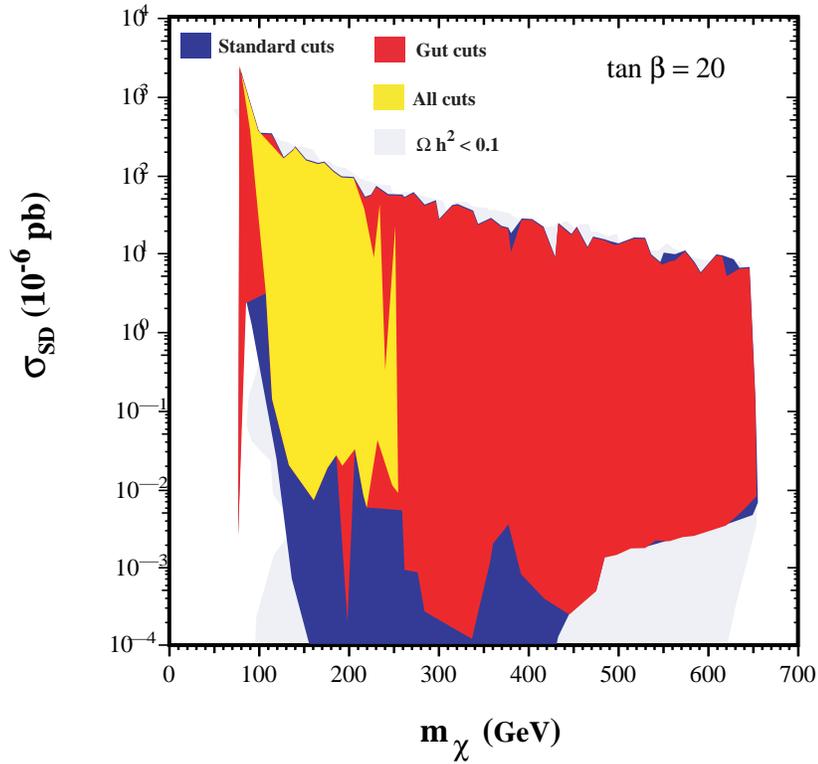,height=4in} \hfill
\end{minipage}
\caption{\label{fig:Andy20}
{\it 
Ranges of (a) the spin-independent and (b) the spin-dependent 
cross sections for $\tan \beta = 20$. The shadings are the same as 
in Fig.~\ref{fig:Andy}.}} 
\end{figure}

Ranges for the spin-independent and -dependent cross sections for $\tan 
\beta = 20$ are shown in Fig.~\ref{fig:Andy20}. Looking first at the 
spin-independent cross section in panel (a), we see that our standard cuts 
on $m_h, \ohsq$ and $b \to s \gamma$ would allow somewhat larger values 
than for $\tan \beta = 10$. This difference is less marked when the GUT 
stability cut is also applied, except for some exceptional parameter 
choices at small $m_\chi$. The jaggedness of the peaks is more
pronounced at this value of $\tan \beta$. Once again, we emphasize that
while we do not expect the area between the peaks to be
empty, the density of points there is
extremely low. When one keeps the low-relic-density points (rescaled
appropriately), we see that indeed the crags are filled in to some extent.
For $\tan
\beta = 20$ these points do not enhance the cross section significantly.
The values of
$m_\chi$ allowed by
$g_\mu - 2$  are larger for $\tan \beta = 20$ than for $\tan \beta = 10$,
and lower  cross sections are also attainable. Overall, the
spin-independent cross  section may vary between $\sim 3 \times
10^{-7}$~pb and
$~10^{-10}$~pb  when $\tan \beta = 20$. 

In the case of the spin-dependent cross section shown in
Fig.~\ref{fig:Andy20}(b), the upper and lower bounds for $\tan \beta = 20$
are very similar to those for $\tan \beta = 10$ if only the standard and 
GUT cuts are applied. However, slightly lower values of the cross section 
are allowed when the $g_\mu - 2$ constraint is applied, without a 
strong dependence on the value of $m_\chi$ (which may be somewhat larger 
than in the case of $\tan \beta = 10$). For $\tan \beta = 20$, the 
spin-dependent cross section ranges between $\sim 2 \times 10^{-3}$~pb and 
$\sim 10^{-8}$~pb.

\begin{figure}
\vspace*{-0.75in}
\hspace*{.70in}
\begin{minipage}{8in}
\epsfig{file=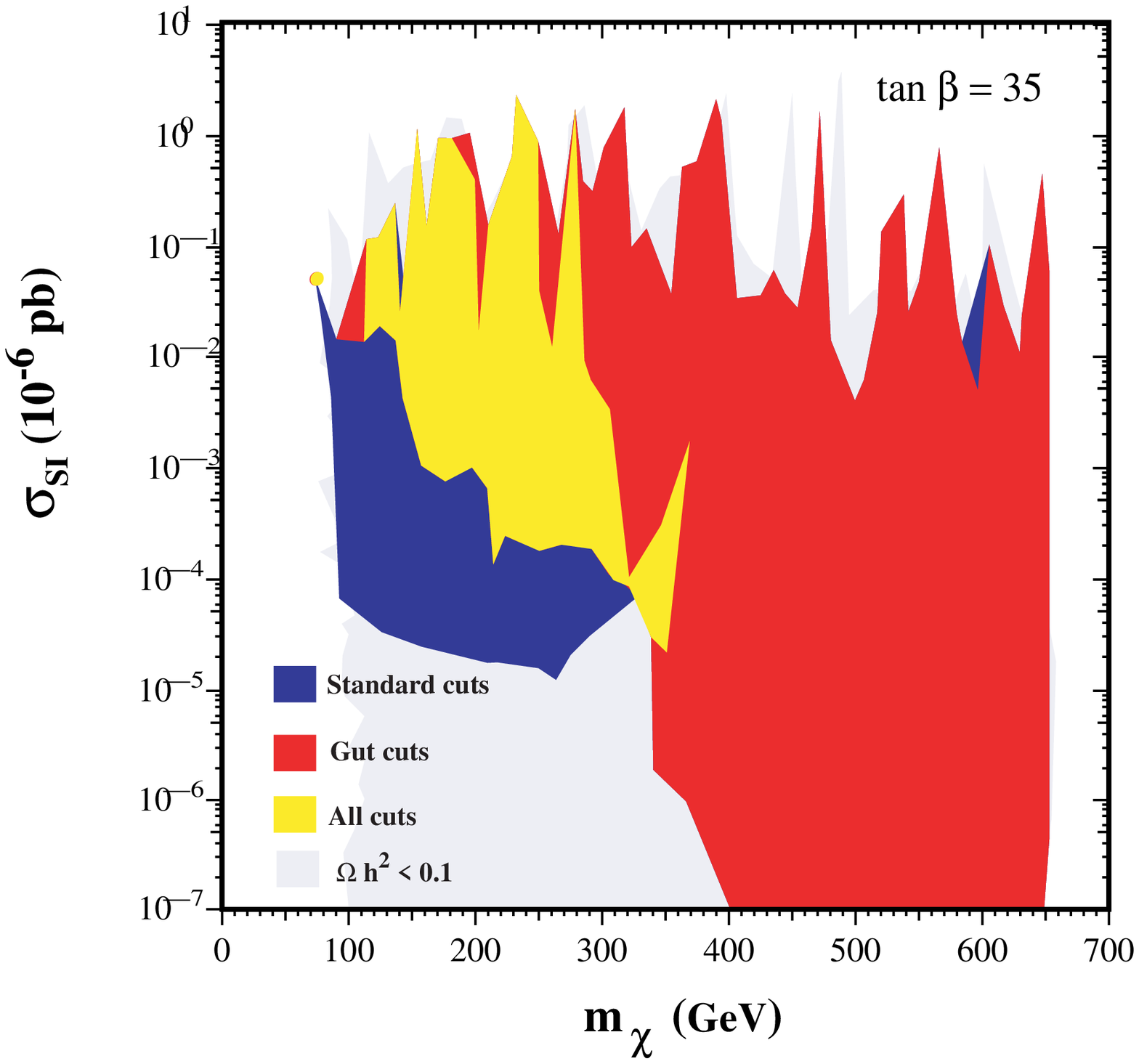,height=4in}
\epsfig{file=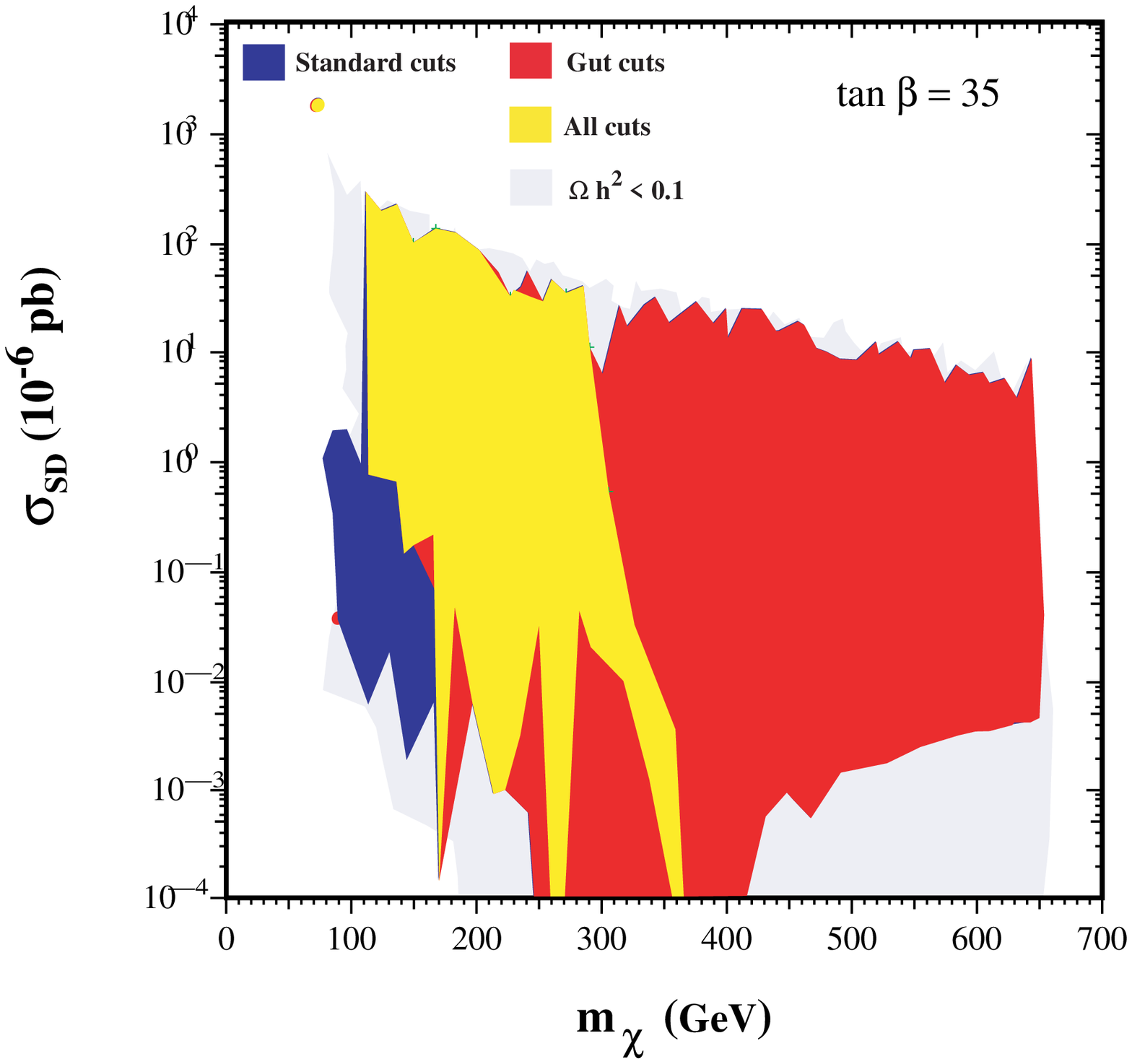,height=4in} \hfill
\end{minipage}
\caption{\label{fig:Andy35}
{\it 
Ranges of (a) the spin-independent and (b) the spin-dependent 
cross sections for $\tan \beta = 35$. The shadings are the same as 
in Fig.~\ref{fig:Andy}.}} 
\end{figure}

\begin{figure}
\vspace*{-0.75in}
\hspace*{.70in}
\begin{minipage}{8in}
\epsfig{file=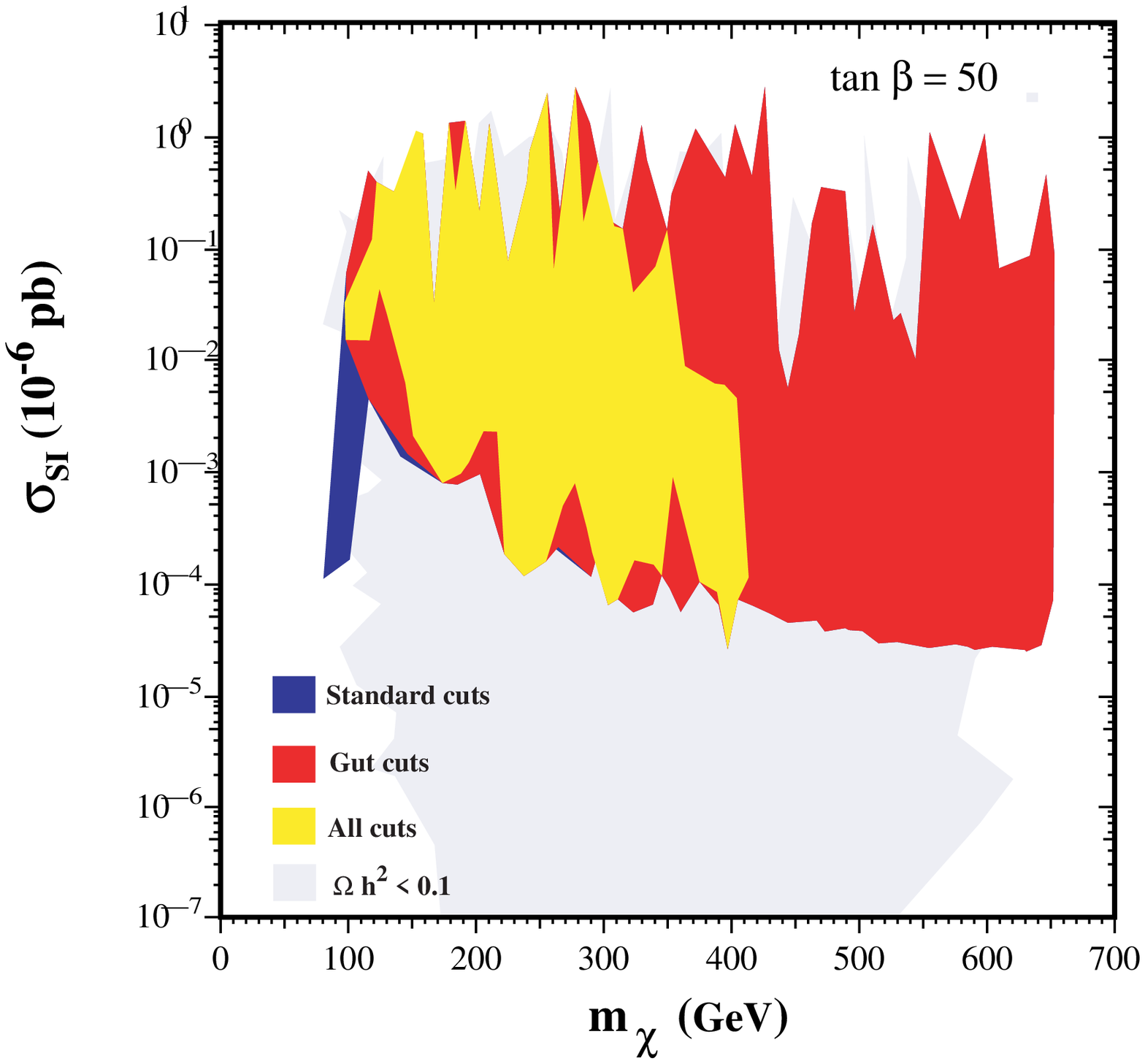,height=4in}
\epsfig{file=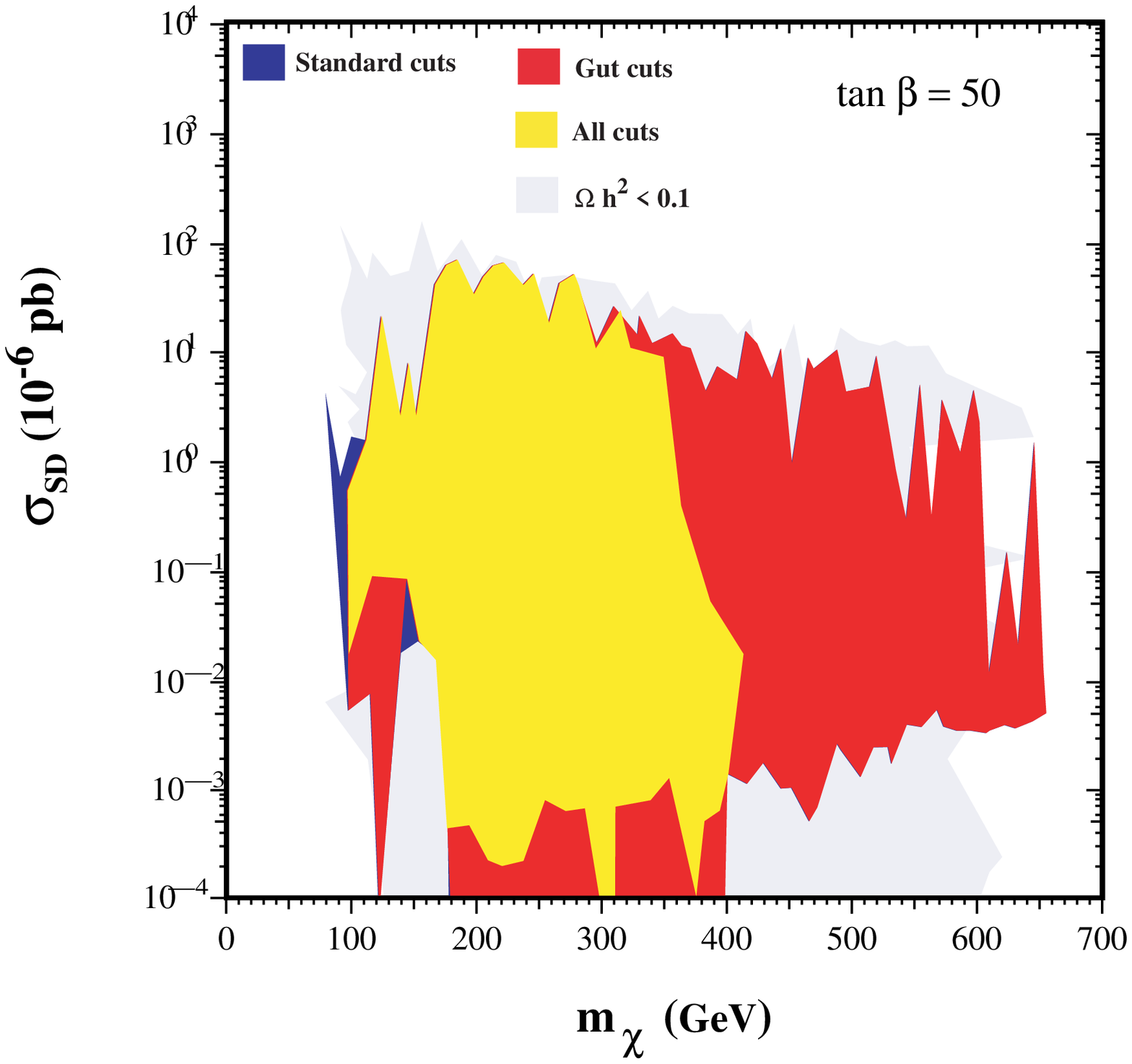,height=4in} \hfill
\end{minipage}
\caption{\label{fig:Andy50}
{\it 
Ranges of (a) the spin-independent and (b) the spin-dependent 
cross sections for $\tan \beta = 50$. The shadings are the same as 
in Fig.~\ref{fig:Andy}.}} 
\end{figure}

Continuing now to $\tan \beta = 35$, as shown in Fig.~\ref{fig:Andy35}, we
see again that the spin-independent cross section may be somewhat larger
still than for $\tan \beta = 20$, though the same is not true for the
spin-dependent cross sections. Once again, the $g_\mu - 2$ constraint
allows larger values of $m_\chi$ as $\tan \beta$ is increased, opening
up the possibility of a smaller cross section, particularly in the
spin-dependent case where a cancellation may occur, potentially
suppressing the cross section by a couple of orders of magnitude. Both
these tendencies are accentuated in the case $\tan \beta = 50$, as shown
in Fig.~\ref{fig:Andy50}. For both $\tan \beta = 35, 50$, the 
spin-independent cross section may, in some exceptional cases, rise above 
$10^{-6}$~pb, even after implementing all the cuts. It may also drop as 
low as $10^{-10}$~pb. The maximal spin-dependent cross section is above 
$10^{-4}$~pb in the $\tan \beta = 35$ case, and somewhat below 
$10^{-4}$~pb in the $\tan \beta = 50$ case. In both cases, it may also 
drop as low as $10^{-10}$~pb.

\begin{figure}
\vspace*{-0.75in}
\hspace*{.70in}
\begin{minipage}{8in}
\epsfig{file=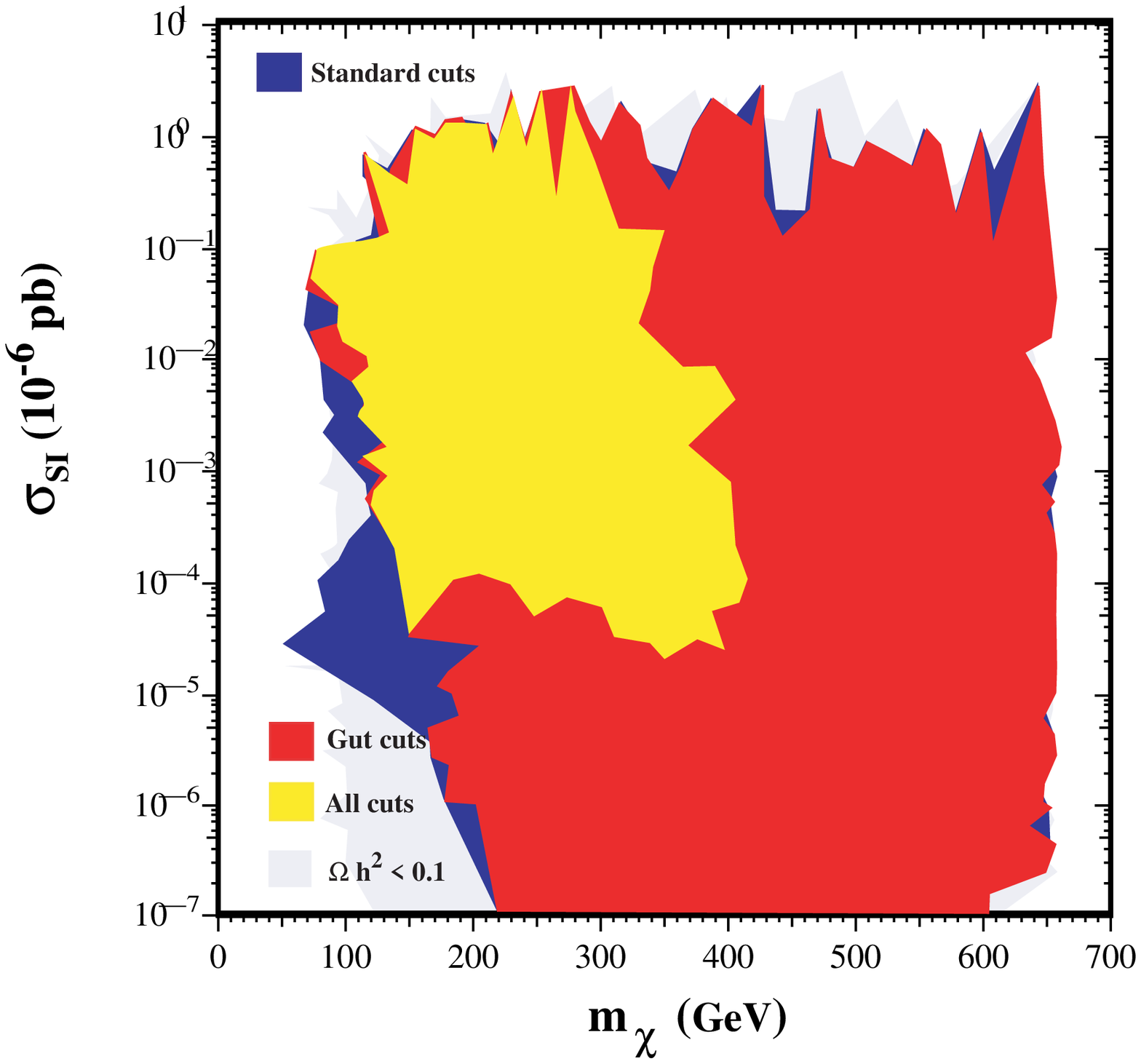,height=4in}
\epsfig{file=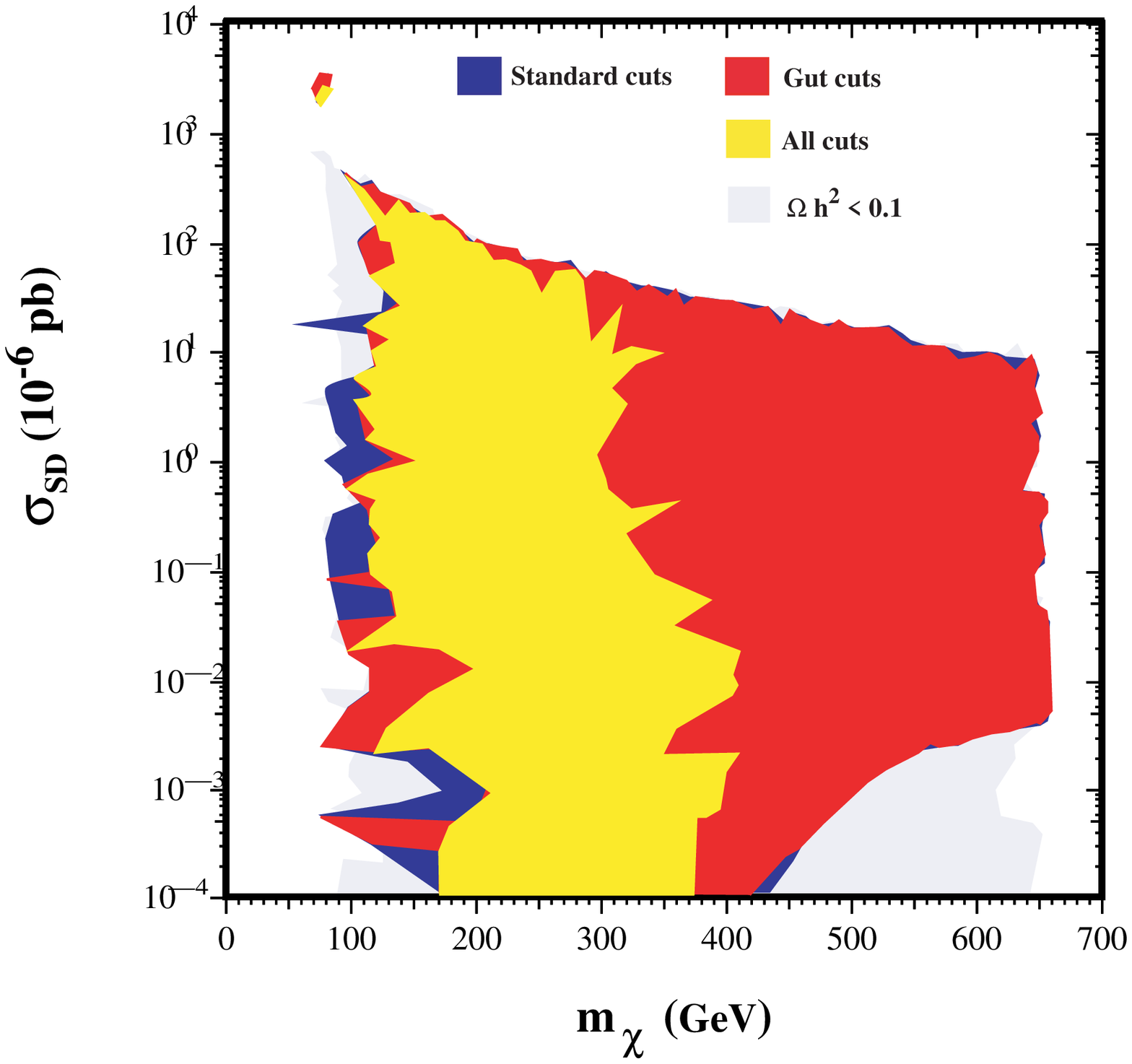,height=4in} \hfill
\end{minipage}
\caption{\label{fig:Andyall}
{\it 
Ranges of (a) the spin-independent and (b) the spin-dependent 
cross sections, sampling randomly all allowed values of $\tan \beta$. 
The shadings are the same as in Fig.~\ref{fig:Andy}.}} 
\end{figure}

Finally, in Fig.~\ref{fig:Andyall}, we display the allowed ranges of (a)
the spin-independent and (b) the spin-dependent cross sections when we
sample randomly $\tan \beta$ as well as the other NUHM parameters. We see
no big surprises compared with the previous plots for individual values of
$\tan \beta$, but observe that the boundaries of the shaded regions are
very ragged, reflecting the finite sample size. After incorporating all
the cuts, including that motivated by $g_\mu - 2$, we find that the
spin-independent cross section has the range $10^{-6}$~pb $\gappeq
\sigma_{SI} \gappeq 10^{-10}$~pb, and the spin-dependent cross section has
the range $10^{-4}$~pb $\gappeq \sigma_{SD} \gappeq 10^{-10}$~pb, with
somewhat larger (smaller) values being possible in exceptional cases. If
the $g_\mu - 2$ cut is removed, the upper limits on the cross sections are
unchanged, but much lower values become possible: $\sigma_{SI} \ll
10^{-13}$~pb and $\sigma_{SD} \ll 10^{-10}$~pb.

\section{Conclusions}

We have discussed in this paper the possible ranges of spin-(in)dependent
elastic cross sections in the NUHM, in which the input soft
supersymmetry-breaking masses of the Higgs doublets are allowed to differ
from those of the squarks and sleptons - which are still assumed to be
universal. Fig.~\ref{fig:Andyall} summarizes the results, including the
flexibility of varying $\tan \beta$ as well as the input scalar and
fermion masses $m_{0,1/2}$, the Higgs mixing parameter $\mu$ and the the
pseudoscalar Higgs mass $m_A$. In this paper, we have not varied the
trilinear soft supersymmetry-breaking parameter $A$, whose effective
low-energy value has in any case a limited range when renormalized from
the GUT scale downwards, so that it does not have a large effect on
the cross sections we study.

We have stressed in this paper the importance of incorporating
consistently all the available phenomenological constraints from
laboratory experiments and cosmology. We have also stressed the importance
of taking into account the running of the NUHM parameters over the full
range of scales between the GUT scale and the electroweak scale. As we
have discussed explicitly, the requirement that the effective scalar
potential be stable at the GUT scale restricts the allowed variations in
the non-universalities of the soft supersymmetry-breaking Higgs masses.

The effects of our phenomenological cuts and this GUT stability
requirement can be seen in Fig.~\ref{fig:Andyall}, and also in the
previous Figs.~\ref{fig:Andy}, \ref{fig:Andy20}, \ref{fig:Andy35} and
\ref{fig:Andy50} for $\tan \beta = 10, 20, 35$ and 50, respectively. Some
examples of our analysis for specific slices through the NUHM parameter
space can be seen in earlier figures.

In general, we find that cross sections may differ by a few orders of
magnitude from those found in the CMSSM, in which the soft
supersymmetry-breaking Higgs masses are assumed to be universal with the
slepton and squark masses at the GUT scale. However, the spin-independent
cross section normally lies well below the present experimental
sensitivity. Only in a few exceptional cases do we find a cross section as
large as the present experimental sensitivity $\sigma_{SI} \sim
10^{-6}$~pb, and a sensitivity $\sigma_{SI} \sim 10^{-10}$~pb would be
required to cover most of the preferred domain of NUHM parameter space.
Even this sensitivity would be insufficient if one disregards the
indication from $g_\mu - 2$, which is the only constraint that motivates a
useful upper bound on the sparticle mass scale, and hence a useful lower
bound on $\sigma_{SI}$. In the case of the spin-dependent cross section,
values of $\sigma_{SD}$ as low as $\sim 10^{-10}$~pb cannot be excluded
even if one takes seriously the $g_\mu - 2$ constraint.

The next logical step in the exploration of the MSSM, relaxing further the
assumption of full scalar-mass universality as in the CMSSM, is to allow
the soft supersymmetry-breaking slepton and squark masses to differ at the
GUT scale. The allowed ranges of the effective low-energy slepton and
squark masses, after renormalization, will be restricted by analogues of
the GUT stability constraints we have applied in this paper. In
particular, we note that general choices of the effective low-energy
slepton and squark masses may lead (in particular) to tachyonic squarks
below the GUT scale, when renormalized to higher scales. We will explore
in future publications the effects on the parameter space and the ranges
of cross sections of applying consistently the GUT stability constraints
to the general non-universal MSSM.

\vskip 0.5in
\vbox{
\noindent{ {\bf Acknowledgments} } \\
\noindent 
The work of K.A.O. and Y.S. was supported in part by DOE grant
DE--FG02--94ER--40823.}

\end{document}